\documentclass[aps,prx,superscriptaddress,twocolumn]{revtex4-2}
\usepackage{graphicx}
\usepackage{xcolor}
\usepackage{amsmath,amssymb}
\usepackage{physics}
\usepackage[utf8]{inputenc}

\newcommand{\br}{\mathbf{r}}
\newcommand{\ba}{\mathbf{a}}
\newcommand{\bA}{\mathbf{A}}
\newcommand{\bK}{\mathbf{K}}

\newcommand{\bk}{\mathbf{k}}
\newcommand{\bG}{\mathbf{G}}
\newcommand{\bg}{\mathbf{g}}

\newcommand{\bq}{\mathbf{q}}
\newcommand{\bE}{\mathbf{E}}

\newcommand{\bB}{\mathbf{B}}
\newcommand{\bv}{\mathbf{v}}
\newcommand{\bj}{\mathbf{j}}
\newcommand{\bu}{\mathbf{u}}
\newcommand{\bL}{\mathbf{L}}

\usepackage{hyperref}
\hypersetup{colorlinks,
linkcolor={blue},
citecolor={blue},
urlcolor={purple}}  

\begin{document}
\title{Unusual magnetotransport in twisted bilayer graphene from strain-induced open Fermi surfaces}

\author{Xiaoyu Wang}
\thanks{These two authors contributed equally}
\affiliation{National High Magnetic Field Laboratory, Tallahassee, Florida, 32310, USA}
\author{Joe Finney}
\thanks{These two authors contributed equally}
\affiliation{Department of Physics, Stanford University, Stanford, CA 94305}
\affiliation{Stanford Institute for Materials and Energy Sciences, SLAC National Accelerator Laboratory, Menlo Park, CA 94025}
\author{Aaron L. Sharpe}
\affiliation{Materials Physics Department, Sandia National Laboratories, Livermore, CA, USA}
\author{Linsey K. Rodenbach}
\affiliation{Department of Physics, Stanford University, Stanford, CA 94305}
\affiliation{Stanford Institute for Materials and Energy Sciences, SLAC National Accelerator Laboratory, Menlo Park, CA 94025}
\author{Connie L. Hsueh}
\affiliation{Department of Applied Physics, Stanford University, Stanford, CA 94305}
\affiliation{Stanford Institute for Materials and Energy Sciences, SLAC National Accelerator Laboratory, Menlo Park, CA 94025}
\author{Kenji Watanabe}
\affiliation{Research Center for Functional Materials, National Institute for Materials Science, 1-1 Namiki, Tsukuba 305-0044, Japan}
\author{Takashi Taniguchi}
\affiliation{International Center for Materials Nanoarchitectonics, National Institute for Materials Science,  1-1 Namiki, Tsukuba 305-0044, Japan}
\author{M. A. Kastner}
\email{mkastner@mit.edu}
\affiliation{Department of Physics, Stanford University, Stanford, CA 94305}
\affiliation{Stanford Institute for Materials and Energy Sciences, SLAC National Accelerator Laboratory, Menlo Park, CA 94025}
\affiliation{Department of Physics, Massachusetts Institute of Technology, Cambridge, MA 02139}
\author{Oskar Vafek}
\email{vafek@magnet.fsu.edu}
\affiliation{National High Magnetic Field Laboratory, Tallahassee, Florida, 32310, USA}
\affiliation{Department of Physics, Florida State University, Tallahassee, Florida 32306, USA}
\author{David Goldhaber-Gordon}
\email{goldhaber-gordon@stanford.edu}
\affiliation{Department of Physics, Stanford University, Stanford, CA 94305}
\affiliation{Stanford Institute for Materials and Energy Sciences, SLAC National Accelerator Laboratory, Menlo Park, CA 94025}
	
\begin{abstract}
    Anisotropic hopping in a toy Hofstadter model was recently invoked to explain a rich and surprising Landau spectrum measured in twisted bilayer graphene away from the magic angle. Suspecting that such anisotropy could arise from unintended uniaxial strain, we extend the Bistritzer-MacDonald model to include uniaxial heterostrain. We find that such strain strongly influences band structure, shifting the three otherwise-degenerate van Hove points to different energies. Coupled to a Boltzmann magnetotransport calculation, this reproduces previously-unexplained non-saturating $B^2$ magnetoresistance over broad ranges of density near filling $\nu=\pm 2$, and predicts subtler features that had not been noticed in the experimental data. In contrast to these distinctive signatures in longitudinal resistivity, the Hall coefficient is barely influenced by strain, to the extent that it still shows a single sign change on each side of the charge neutrality point -- surprisingly, this sign change no longer occurs at a van Hove point. The theory also predicts a marked rotation of the electrical transport principal axes as a function of filling even for fixed strain and for rigid bands. More careful examination of interaction-induced nematic order versus strain effects in twisted bilayer graphene could thus be in order.
\end{abstract}
\maketitle

\section{Introduction}

\begin{figure*}[ht]
    \centering
    \includegraphics[width=\linewidth]{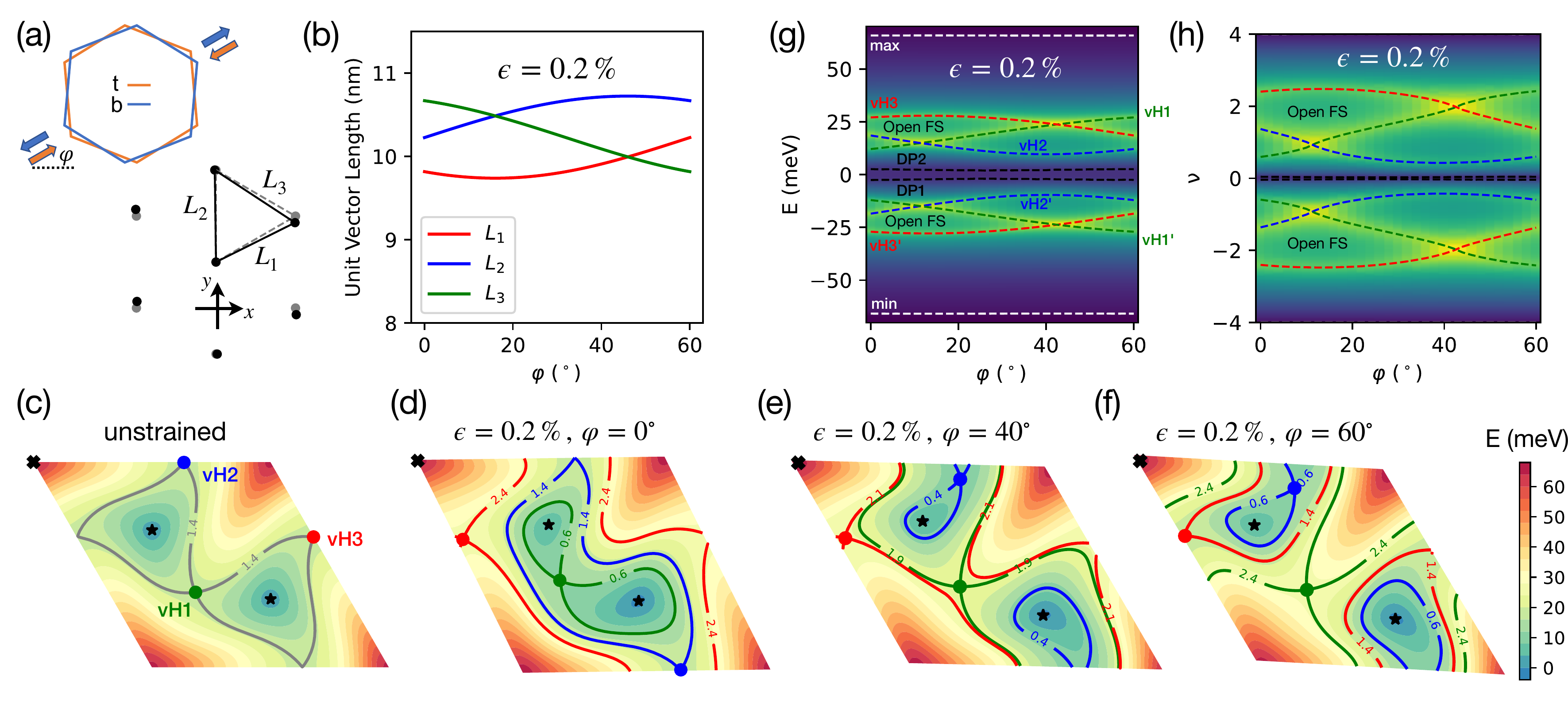}
    \caption{{\bf(a)} Schematics of applying a uniaxial heterostrain on the pair of  microscopic unit cells of monolayer graphene making up TBG. (Upper sketch) Orange (blue) color corresponds to top (bottom) layer. The uniaxial strain of strength $+ (-) \epsilon/2$ and direction $\varphi$ on the top (bottom) layer are represented as colored arrows. (Lower sketch) Deformation of the moir\'e superlattice for twist angle $1.38^\circ$ due to a uniaxial heterostrain of $\epsilon= 0.2\%$ and $\varphi=0^\circ$. Unstrained (gray, dashed) and strained (black, solid) triangular lattice sites of AA stacking regions of the moir\'e superlattice are depicted. {\bf(b)} Dependence of the three moir\'e triangular bond lengths on $\varphi$ for a fixed strength. {\bf(c-f)} Energy maps of the upper band of the BM Hamiltonian in valley $\bK$, plotted in the moir\'e Brillouin zone specified by $\bk=k_1\bg_1+k_2\bg_2$, where $k_{1,2}\in[0,1)$. There are six special points of the band structure, i.e., two Dirac points (black stars), three van Hove points (colored dots), and one band maximum (black cross). The contour lines intersecting the van Hove points are plotted and labeled by their respective filling fractions. In the unstrained case (c), the two Dirac points and three van Hove points are respectively at equal energies. The energy degeneracies are lifted in the presence of uniaxial heterostrain, as illustrated in (d-f). This leads to semimetallic behavior at the CNP, and a $\varphi$-dependent filling range near $\nu=2$ with open FSs. {\bf(g-h)} $\varphi$-dependence of the energies and filling fractions of the band structure special points for a fixed heterostrain strength. The background colormap is the calculated density of states, with a broadening of $\delta=1\mathrm{meV}$. Green (blue) color represents high (low) density of states. The energetic minimum and maximum of the narrow bands are shown with horizontal dashed grey lines.}
    \label{fig:strain_energetics}
\end{figure*}

The discovery of superconductivity and correlated insulating states in magic-angle twisted bilayer graphene (TBG) \cite{Cao2018a,Cao2018b} placed the material at the forefront of condensed matter physics research \cite{Kerelsky2019,Lu2019,Jiang2019,Yankowitz2019,Choi2019,Sharpe2019,Xie2019,Zondiner2020,Wong2020,Serlin2020,Stepanov2020,Cao2020nematicity,Liu2021,Yacoby2021,Wu2021}. The moir\'e superlattice potential of TBG, resulting from a small relative twist angle $\theta$ between the graphene layers, can induce nearly flat, topologically non-trivial, isolated bands, consisting of electronic states near the Dirac points of each monolayer of graphene \cite{BM2011}. As a result, TBG is an exceptional platform for studying the interplay of electron correlations and band topology \cite{Andrei2020,Balents2020,Koshino2018,Po2018,Kang2018,Kang2019,Zhida2019,Xie2020,Bultinck2020b,ZhangYi2020,Cea2020,Kang2020a,Kang2020b,TBGIV2020,TBGV2020,TBGVI2020,Potasz2021,Kwan2021,Parker2021}.

Strain -- especially heterostrain consisting of differing lattice distortions in the two layers -- is believed to play an important role in the phase diagram of TBG \cite{ZhenBi2019,Parker2021,Kwan2021}. Scanning probe measurements typically find uniaxial heterostrain in the range of $0.1 - 0.7\%$ in samples fabricated with the tear-and-stack method \cite{Kerelsky2019,Xie2019,Wong2020}. For heterostrain, as opposed to homostrain, the linear distortion of the moir\'e unit cell is amplified by a factor of $\sim1/\theta$ relative to the linear distortion of the microscopic atomic lattice. Because we infer twist angle from moir\'e unit cell area in transport, this effect leads to underestimates of the uncertainty in twist angles presented in transport literature, as noted in Ref. \cite{Kerelsky2019}. For example, $0.2\%$ uniaxial heterostrain causes a $\sim 8\%$ change in the linear size of the moir\'e unit cell for a twist angle of $1.38^\circ$. However, the effect on the moir\'e unit cell area is much reduced.

In a recent report by some of the authors [Finney \textit{et al}, Ref.~\cite{FinneyPNAS2022}], a TBG sample with a moir\'e unit cell area of 90 nm$^2$ (corresponding to $\theta=1.38^\circ$, well above the magic angle) displayed several unusual phenomena in magnetotransport. The sample did not exhibit the strong interaction driven effects typically observed in near-magic-angle devices. Rather, over a broad filling range near half filling, the longitudinal magnetoresistivity (MR) exhibited a $B^2$ increase up to $\approx 5$ T, after which quantum oscillations set in. Such $\sim100$-fold increase in MR was not explained, although the authors conjectured that strain may have played a role based on comparison of a toy Hofstadter model with anisotropy, over a broader range of magnetic field.

In this work, we present a systematic theoretical study of the impact of uniaxial heterostrain on the narrow-band dispersion of TBG above the magic angle, analyze its consequences for weak field magnetotransport, and compare it with experimental data from Ref.~\cite{FinneyPNAS2022}. We base our theory on the Bistritzer-MacDonald (BM) continuum model \cite{BM2011}, incorporating heterostrain in the form of a deformation potential, a pseudo-magnetic field \cite{Ando2002,NamKoshino2017}, and a distortion of the moir\'e pattern in the interlayer tunneling.

Our key theoretical result is that heterostrain lifts the energetic degeneracy of the two Dirac points as well as that of the three van Hove points of a given band. The splitting of the two Dirac points leads to a semimetallic state near the charge neurality point (CNP) with small Fermi pockets. More interestingly, the splitting of the van Hove points leads to open Fermi surfaces (FS) in the filling range bounded by two of the van Hove points. In the weak field semiclassical regime governed by the Boltzmann equation, the open FSs generally lead to a non-saturating $B^2$ MR, explaining the low-field experimental findings of Ref.~\cite{FinneyPNAS2022}. 

This theory makes a number of falsifiable predictions. Of note, it predicts a large degree of mixing between longitudinal and transverse MRs within the open FS regime, due to an uncontrolled misalignment of the strain-induced principle axis of transport and the direction of current flow in the Hall bar. It predicts a subtle cusp in resistivity corresponding to the crossing of the lowest-energy van Hove point. Finally, it predicts a Lifshitz transition from two FS pockets to one upon crossing this lower van Hove point. We reanalyze experimental data from \cite{FinneyPNAS2022}, and find that these predictions are verified. The theory does not capture the electron-hole asymmetry in the experimental data.

The theory also has a few unexpected features. Firstly, the sign change singularity in the Hall number, one on each side of CNP, does not coincide with any of the van Hove points and instead occurs inside the filling range with open FSs. Secondly, the transport principal axis continuously rotates by up to $90^\circ$ as density is tuned from the CNP to the open FS regime. Such rotation of the transport axes is generally associated with interaction-induced nematic order \cite{Cao2020nematicity}, but here we find that it can arise purely due to strain-induced band structure effects.

This work clearly demonstrates that the effects of even miniscule amounts of heterostrain in TBG cannot be neglected. Dramatic and unexpected phenomena occur in strained TBG even in the single-particle regime, without the strong correlation effects that arise near the magic angle. Given the amplifying effect of a small heterostrain on the moir\'e length scale, it is tantalizing to consider strain engineering of such devices to achieve effects that would be impossible in regular solids due to structural instabilities.

\section{Geometric and energetic effects of uniaxial heterostrain on TBG} 

In the limit of small deformations, both the uniaxial heterostrain and a small twist angle are captured via a coordinate transformation: $\br'_l = \br + \bu_l(\br)$, where $l=t,b$ labels the top (bottom) graphene layers, and $\bu_l(\br) \approx \mathcal{E}_l \br$ is the local deformation field. The symmetric and antisymmetric part of the $2\times 2$ tensor $\mathcal{E}_l$ describes strain and rotation respectively. For twist angle $(\theta)$ and a uniaxial heterostrain of strength $(\epsilon)$ and direction $(\varphi)$, we parameterize $\mathcal{E}_{t}=-\mathcal{E}_{b}\equiv \mathcal{E}/2$, where $\mathcal{E}\equiv \mathcal{T}(\theta)+ \mathcal{S}(\epsilon,\varphi)$, and given by: 
\begin{equation}
\mathcal{T}(\theta) = \begin{pmatrix} 0 & -\theta\\ \theta & 0\end{pmatrix},\ \mathcal{S}(\epsilon,\varphi) = R_{\varphi}^T \begin{pmatrix} -\epsilon & 0\\ 0 & \nu\epsilon \end{pmatrix}R_{\varphi}.
\end{equation}
Here $R_\varphi$ is the two-dimensional rotation matrix, and $\nu\approx 0.16$ is the Poisson ratio  \cite{Kerelsky2019}. Physically, $\epsilon>0$ corresponds to compressing the top layer while streching the bottom layer along the direction determined by $\varphi$, as illustrated in Fig.~\ref{fig:strain_energetics}(a). A relative deformation $\mathcal{E}$ between the graphene bilayers generates a moir\'e superlattice, with moir\'e reciprocal lattice vectors $\bg_{i=1,2}=\mathcal{E}^T\bG_{i=1,2}$, where $\bG_{i}$ are reciprocal lattice vectors of the undeformed monolayer graphene. The moir\'e  lattice vectors $\bL_{i=1,2}$ are uniquely defined through the relation $\bL_i\cdot \bg_j=2\pi \delta_{ij}$. It is important to note that only relative deformations generate the moir\'e superlattice. Homogenous deformations do not play an important role in the narrow band physics, and we neglect it in this work \footnote{We checked numerically that adding a small homogeneous strain in addition to a heterostrain of similar strength yields almost identical band and transport properties to the case of adding a heterostrain alone.}.

Under rotation $R_\varphi$, the strain tensor transforms as a headless vector that remains invariant under $\varphi\rightarrow \varphi+180^\circ$. Combined with the $C_{3z}$ symmetry of the undeformed graphene lattice, the strained electronic dispersion within a given graphene valley simply rotates $60^\circ$ under $\varphi\rightarrow \varphi + 60^\circ$. We hereby will only report results for $\varphi\in[0^\circ,60^\circ)$. For concreteness we define the microscopic unit cell vectors $\ba_{i=1,2}$ of undeformed graphene lattice as $ \ba_1 = a(\frac{1}{2},-\frac{\sqrt{3}}{2}),\ \ba_2 = a(1,0)$, where $a\approx2.46\AA$ is the lattice constant. The positions of the sublattice A,B within a unit cell are chosen as $\vec{\tau}_A = (0,0)$ and $\vec{\tau}_B = \frac{a}{\sqrt{3}}(0,1)$. The reciprocal lattice vectors are $\bG_1 = \frac{4\pi}{\sqrt{3}a}(0,-1)$ and $\bG_2 = \frac{4\pi}{\sqrt{3}a}(\frac{\sqrt{3}}{2},\frac{1}{2})$. Different conventions lead to different definitions of the Dirac Hamiltonian (see for instance Ref.~\cite{ZhenBi2019}), but the physics is consistent. 

Fig.~\ref{fig:strain_energetics}(a-b) illustrates the geometric effects of heterostrain for twist angle $\theta=1.38^\circ$. For $\epsilon=0.2\%$, typical in these systems \cite{Kerelsky2019,Xie2019,Wong2020}, there is a large change in the bond length of the neighboring AA-stacked regions ($\bL_{i=1,2,3}$) of the moir\'e triangular superlattice, which used to form an equilateral triangle at $\epsilon=0$. The effect of heterostrain on the moir\'e unit cell vectors can be estimated to be as large as $\epsilon/\theta\approx 8\%$. However, the effect on the moir\'e unit cell area is much smaller at $\nu^2\epsilon^2/\theta^2$ (see Supplementary Material (SM) Sec. I). Such dramatic amplification of the microscopic strain makes moir\'e materials ideal for strain engineering not achievable in conventional materials due to structural instability. 

We proceed to discuss the energetic effects in the context of the continuum BM model \cite{BM2011}. We work in the limit where both $\mathcal{E}_l$ and the wavevector $\bk$ in the moir\'e Brillouin zone are small, and consider only the leading order terms in both. This would mean, for instance, that terms such as $\mathcal{E}\bk$ are omitted as higher order terms. This treatment is generally justified away from the magic angle, because higher order terms can play an important role only close to the magic angle where the narrow-band bandwidth is suppressed to a similar energy scale \cite{Vafek2022a,Kang2022b}. Furthermore, we checked that at $\theta\approx 1.38^\circ$ the effects of such higher order terms are indeed negligibly small. To leading order, the strained BM Hamiltonian for a given valley is given by:
\begin{equation}\label{eq:strained_bm}
    H_{\eta} = (\sum_{l=t,b}  H_{\eta,l}^{intra}) + H_{\eta}^{inter} ,
\end{equation}
where $\eta=\pm 1$ labels $\bK\ (\bK')$ valleys of monolayer graphene. The interlayer Hamiltonian is given by:
\begin{equation}
    H_{\eta,l}^{inter}  \approx \int \mathrm{d}^2\br \psi^\dagger_{\eta,t} \left(\sum_{j=1,2,3} T_{\eta,j} e^{-i\eta \bq_{j}\cdot\br}\right)\psi_{\eta,b}(\br) + h.c.,
\end{equation}
where $\psi_{\eta,l}(\br)\equiv (\psi_{\eta,l,A}(\br),\psi_{\eta,l,B}(\br))^T$ is a spinor in the sublattice basis for a given valley and layer. We have suppressed the spin index for simplicity. $\bq_{j=1,2,3}$ are the three nearest neighbor bonds of the reciprocal honeycomb lattice, and
\begin{equation}
    T_{\eta,j} = w_0\sigma_0+w_1\left(\cos \frac{2\pi(j-1)}{3} \sigma_x+ \eta \sin \frac{2\pi(j-1)}{3}\sigma_y\right).
\end{equation}
$(\sigma_0,\sigma_x,\sigma_y)$ are Pauli matrices acting on sublattice degrees of freedom. 

The intra-layer Hamiltonian is given by:
\begin{equation}
\begin{split}
     & H_{\eta,l}^{intra} =  \alpha \sum_{\bk} \psi^\dagger_{\eta,l}(\br)  ( \tr[\mathcal{E}_l] \sigma_0)\psi_{\eta,l}(\br)\\
     &-  \frac{\hbar v_F}{a}\sum_{\bk} \psi^\dagger_{\eta,l}(\br)  \left[ (\bk - \bA_{\eta,l})\cdot(\eta \sigma_x,\sigma_y) \right]\psi_{\eta,l}(\br).
\end{split}
\end{equation}
Here the first term is the deformation potential that couples to the electron density. Its value is not precisely known in the literature, with numbers ranging from $-4.1\ \mathrm{eV}$  to $30\ \mathrm{eV}$ depending on the methodology \cite{Hwang2008,Efetov2010,Karsten2012,Grassano2020}. We use $\alpha=-4.1\ \mathrm{eV}$ in this work based on first principles calculations \cite{Grassano2020}, although the deformation potential does not have an important effect on the band dispersions for heterostrain $\epsilon\approx 0.2\%$, and only leads to minor quantitative differences.  $\bA_{\eta,l}$ is the pseudovector potential that comes from changes in the inter-sublattice hopping due to deformations, and changes sign between graphene valleys. It is given as \cite{Ando2002,NamKoshino2017}: $\bA_{\eta,l} = \frac{\sqrt{3}\beta}{2a}\eta (\epsilon_{l,xx}-\epsilon_{l,yy},-2\epsilon_{l,xy})$, where we choose $\beta \approx 3.14$ from Refs.~\cite{Kerelsky2019,ZhenBi2019}. We shall further fix $\hbar v_F/a=2.68\mathrm{eV}$, $w_0=88\mathrm{meV}$, and $w_1=110\mathrm{meV}$ in our calculations, and also set $\hbar=1$ in the remainder of the paper.

To leading order approximation, the strained BM Hamiltonian in a given valley (Eq.~(\ref{eq:strained_bm})) has particle-hole symmetry under $P \psi_{l}(\br) = \sum_{l'}i(\mu_y)_{ll'}\psi_{l'}(-\br)$ \cite{TBGI2020}, where $\mu_y$ is a Pauli matrix acting on the layer degrees of freedom. This means that for every single electron state at energy $E$ and wavevector $\bk$, there is a state at energy $-E$ and wavevector $-\bk$. This particle-hole symmetry has been investigated extensively for the unstrained BM model, e.g., Refs.~\cite{Zhida2019,Kang2021}, and here it is generalized to the strained case. Since in experiments particle-hole asymmetry is evident for the off-magic-angle device \cite{FinneyPNAS2022}, they would come from either higher order gradient terms beyond what's captured in the BM model in Eq.~(\ref{eq:strained_bm}), or due to interaction effects \cite{Guinea2018,Rademaker2019,Goodwin2020,Choi2021}, or their combination.

We proceed to discuss the heterostrain effects on the band structure with $\epsilon=0.2\%$ and varying direction specified by $\varphi\in[0^\circ,60^\circ)$, depicted in Fig.~\ref{fig:strain_energetics}(d-f). For simplicity we only show contour maps of the upper band from valley $\bK$ in the moir\'e Brillouin zone specified by $\bk=k_1\bg_1+k_2\bg_2$, where $k_{1,2}\in[0,1)$. Heterostrain preserves $C_2T$ and valley $U(1)$ \cite{Po2018} and therefore the lower and upper bands remain connected via two Dirac points. The upper band features six special points --- two Dirac points (black stars), three van Hove points (colored dots), and one band maximum (black cross).  The six special points of a given band are related to ``critical points" in the context of the Morse theory, which states that
\begin{equation}\label{eq:morse_equality}
    \sum_{i}(-1)^{\gamma_i}=\chi,
\end{equation}
where $\gamma_i$ is the index of the $i$-th critical point, and $\chi$ is the Euler characteristic of a manifold \cite{audin_damian_2014}; $\chi$ vanishes for the Brillouin zone which is a torus. Although a Dirac point is strictly a point of non-analyticity and is not directly covered by Morse theory, if we imagine adding a tiny gap term it will become a legitimate band extremum and Morse theory applies. Whereas the two band minima (Dirac points) and the band maximum have even $\gamma$ and so each contributes $+1$ to the sum, every conventional van Hove point (i.e. not a higher order) has an odd $\gamma$ and contributes $-1$. Their sum thus vanishes. Therefore, the van Hove points can only be annihilated/created by colliding with local minima/maxima. For a relatively small heterostrain as shown in Fig.~\ref{fig:strain_energetics}, the number of special points per band is the same as at $\epsilon=0$. However for larger heterostrain (e.g., $\epsilon=0.5\%$,  see SM Fig.~1), more striking behavior of the special points can occur, such as a change in their total number via afore mentioned collisions and the appearance of tilted type II Dirac cones \cite{Goerbig2008,Soluyanov2015}.

A key finding of the present work is that the respective energy degeneracies of the two Dirac points and the three van Hove points are lifted by uniaxial heterostrain, and depend sensitively on $\varphi$. In the absence of strain [Fig.~\ref{fig:strain_energetics}(c)], the three van Hove points are at equal energy, and separate closed contours of constant energy centered around the Dirac points from closed contours centered around band maximum. As illustrated in Fig.~\ref{fig:strain_energetics}(d-f), uniaxial heterostrain splits the energy degeneracy of the two Dirac points, leading to a semimetallic state with small Fermi pockets near CNP \cite{ZhenBi2019}.  The three van Hove points also split in energy. The two outermost van Hove points (i.e., closer to the band maximum) bound a filling range of open FSs near $\nu=2$, while the innermost van Hove point moves closer to one of the Dirac points. If we continue increasing $\epsilon$, a collision of the critical points occurs, the innermost van Hove disappears, the two Dirac points become type-II tilted, and a new ordinary minimum is created. Note that a small mass added to type-II tilted Dirac points won't introduce band extrema and as a consequence type-II tilted Dirac points are not critical points of Morse theory, therefore after the collision Eq.~(\ref{eq:morse_equality}) still holds. 

Interestingly, the elongation of the FSs shows a strong filling dependence. Close to the CNP, the bigger Fermi pocket that encloses a Dirac point is stretched along a perpendicular direction to that of the open FSs, see Figs.~\ref{fig:strain_energetics}(d-f). As explained later, this leads to a dramatic rotation of the principal transport axis when the filling is tuned from the CNP to the open FS range.  

The dependence of the energy and filling of the band structure special points on $\varphi$ at a fixed $\epsilon$ is shown in Fig.~\ref{fig:strain_energetics}(g-h). Of notable interest is the sensitivity of the filling range with open FSs to $\varphi$. This filling range must in fact vanish at some $\varphi$ between $0^\circ$ and $60^\circ$, when the energies of the two outermost van Hove points cross. As seen in Fig.~\ref{fig:strain_energetics}(d-f), this also alters the elongation of the open FSs.

\section{Boltzmann equation and Magnetoresistivity in TBG}

Having understood the heterostrain effects on the bandstructure, we proceed to discuss the implications for magnetotransport. We begin by considering the general structure of the two-dimensional resistivity tensor ${\rho}$ subject to heterostrain. The resistivity tensor is defined via:
\begin{equation}
    \begin{pmatrix}
    E_x \\ E_y 
    \end{pmatrix} = 
    \begin{pmatrix}
        \rho_{xx} & \rho_{xy}\\
        \rho_{yx} & \rho_{yy}
    \end{pmatrix}
    \begin{pmatrix}
    j_x \\ j_y
    \end{pmatrix},
\end{equation}
where $\bE=(E_x,E_y)^T$ and $\bj=(j_x,j_y)^T$ are electric field and current vectors respectively. Under rotation by $\delta \theta$, the resistivity tensor transform as:
\begin{equation}
    {\rho}' = R_{\delta\theta}^T {\rho} R_{\delta\theta},\ R_{\delta\theta} = \begin{pmatrix}
    \cos \delta\theta & -\sin \delta\theta \\
    \sin \delta\theta & \cos \delta\theta
    \end{pmatrix}.
\end{equation}

If the underlying system has a point group symmetry that is higher than $C_{2z}$ (e.g., $C_{3z},C_{6z}$), then ${\rho}=\rho_0\mathbb{I} - i \rho_H \tau_y$ is the most general form of ${\rho}$ invariant under such rotations. Here $\tau_y$ is the Pauli matrix acting in the two-dimensional coordinate basis, $\rho_0(-B)=\rho_0(B)$ is the longitudinal resistivity, and $\rho_H(-B)=-\rho_H(B)$ is the Hall resistivity. The even/odd parity under time reversal is guaranteed by the Onsager reciprocal relations.

Since heterostrain breaks the point group symmetry down to $C_{2z}$, we generally expect $\rho_{xx}\neq \rho_{yy},\ \rho_{xy}\neq -\rho_{yx}$.
Nevertheless, it is always possible to define transport principal axes after a suitable rotation $\delta\theta$ of the coordinate system, such that: 
\begin{equation} \label{eq:mr_principal}
    {\rho}_{\text{principal}}= \frac{1}{2}(\rho_1+\rho_2)\mathbb{I}+\frac{1}{2}(\rho_1-\rho_2)\tau_z + \rho_H i\tau_y.
\end{equation}
Here $\rho_{1,2}$ are longitudinal resistivities along the principal transport directions $\hat{e}_{1,2}$ respectively. The rotation angle $\delta \theta$ is determined up to $180^\circ$ by requiring $\rho_1< \rho_2$. 

Below we first derive the MR tensor using Boltzmann approach for a general non-interacting electronic system within the relaxation time approximation. Since there is currently insufficient understanding of the scattering mechanisms determining electrical transport in TBG, here we follow Ref.~\cite{Ming2021} and use relaxation time approximation. We will then present the results for heterostrained TBG, showing that in the open FS region, the low resistivity principal axis ($\hat{e}_{1}$) is nearly perfectly aligned with the shortest moir\'e bond direction. However there is a dramatic rotation of the principal axis as the filling moves towards the CNP. We further show that the open FSs lead to a $B^2$ non-saturating MR along $\hat{e}_{2}$, and a saturating resistivity along $\hat{e}_{1}$. For random orientation ($\theta_0$) of the principal axis to the electrical current axis in the Hall bar geometry, e.g.,  as in Ref.~\cite{FinneyPNAS2022}, the longitudinal resistivity is given by: $\rho_{xx} =\rho_1\cos^2 \theta_0+\rho_2\sin^2\theta_0$. It is dominated by the $\rho_2\sim B^2$ component, and as a result, the experimental measurements should observe the non-saturating MR component if there is a misalignment with respect to the principal transport axis.

\subsection{Boltzmann equation and method of characteristics}
We begin with a brief description of the method of characteristics used to solve the Boltzmann equation perturbatively in electric field ${\bf E}$ but without a restriction on the strength of the perpendicular magnetic field ${\bf B}=B\hat{z}$, as long as the semiclassical regime holds \cite{Lifshits1973}. Due to $C_{2z}T$ symmetry of TBG at ${\bf B}=0$, there is no Berry curvature contribution to the semiclassical equations of motion. Then, within the relaxation time approximation, the Boltzmann equation for a given energy band becomes 
\begin{equation}
    \frac{\partial n_\bk}{\partial t} + (q\bE+q\bv_\bk\times \bB) \cdot \frac{\partial n_\bk}{\partial \bk} =  - \frac{n_{\bk}-n_{0,\bk}}{\tau},
\end{equation}
where $q\bE+q\bv_\bk\times \bB$ is the total force on the Bloch electrons, with $\bv_\bk \equiv \nabla_{\bk} \varepsilon_{\bk}$ and charge $q$; $n_{0,\bk}$ is the equilibrium Fermi-Dirac distribution and $n_{\bk}$ is the desired non-equilibrium distribution function.

We consider a stationary solution to the Boltzmann equation by parameterizing the distribution function as:
\begin{equation}
    n_{\bk} = n_{0,\bk} + n_{1,\bk}.
\end{equation}
As a result, the Boltzmann equation for the deviation of the distribution function from equilibrium is:
\begin{equation}
    (q\bE\cdot \bv_{\bk}) \frac{\partial n_{0,\bk}}{\partial \varepsilon_\bk}+(q\bv_\bk\times \bB)\cdot \frac{\partial n_{1,\bk}}{\partial \bk}   =  - \frac{n_{1,\bk}}{\tau}.
\end{equation}
Note that the magnetic field only couples to $n_1$ since $(q\bv_\bk\times\bB)\cdot \nabla_\bk n_{0,\bk}=(q\bv_\bk\times\bB)\cdot\bv_{\bk} \partial_{\varepsilon_{\bk}} n_{0,\bk}=0$. 

In order to solve the above partial differential equation (PDE), we seek a family of curves covering the $\bk$-space which we parameterize
as $\bk(s)$ with $s\in[0,s_0)$, such that along these curves the PDE becomes an ordinary differential equation (ODE).
If a curve $\bk(s)$ satisfies
\begin{equation} \label{eq:trajectory}
    \frac{d\bk(s)}{d s} = q\bv(s)\times \bB,
\end{equation}
then $n_{1,\bk(s)}\equiv n_{1}(s)$ satisfies
\begin{equation}
    (q\bE\cdot \bv_{\bk}) \frac{\partial n_{0,\bk}}{\partial \varepsilon_\bk}|_{\bk=\bk(s)}+ \frac{d n_{1}(s)}{d s}   =  - \frac{n_{1}(s)}{\tau}.
\end{equation}
Because
\begin{equation}
    \frac{d \varepsilon(s)}{d s} =  \bv(s) \cdot \frac{d \bk(s)}{d s} = 0,
\end{equation}
the curve $\bk(s)$ must coincide with the contour of constant energy. Thus, the Boltzmann equation becomes:
\begin{equation}
    [q\bE\cdot \bv(s)] \frac{\partial n_{0}(s)}{\partial \varepsilon(s)}+ \frac{d n_{1}(s)}{d s}   =  - \frac{n_{1}(s)}{\tau}.
\end{equation}
The ODE is readily solved with:
\begin{equation}
    n_{1}(s) = \chi_0 e^{-s/\tau} - e^{-s/\tau}\int_0^{s}\mathrm{d}s' e^{s'/\tau} [q\bE\cdot \bv(s')] \frac{\partial n_{0}(s')}{\partial \varepsilon(s')}.
\end{equation}
where $\chi_0$ is a constant determined by the following argument.
Since $\bk(s)$ describes a constant energy contour in a two-dimensional Brillouin zone, it is either a closed contour, or several open contours that terminate on boundaries of the Brillouin zone such that they form a closed loop on a torus. In either case, $\bk(s)$ is periodic under $s\rightarrow s+s_0$ modulo a moir\'e reciprocal lattice vector, where $s_0$ is the periodicity. The periodicity condition $n_{1}(s_0)=n_{1}(0)$ leads to
\begin{equation}
    \chi_0 = \frac{1}{1-e^{s_0/\tau}}\int_0^{s_0}\mathrm{d}s' e^{s'/\tau}(q\bE\cdot \bv(s')) \frac{\partial n_{0}(s')}{\partial \varepsilon(s')},
\end{equation}
which determines the desired $n_1(s)$.

In the low temperature limit, the steady state current from a given energy band is calculated as:
\begin{equation} \label{eq:current}
\begin{split}
    j^\mu & = q\int\frac{\mathrm{d}^2\bk}{(2\pi)^2} v^{\mu}_{\bk} n_{1,\bk} \\
    & =\frac{q^2B}{(2\pi)^2} \int \mathrm{d} {\varepsilon}\int_0^{s_0} \mathrm{d} s v^{\mu}(s) n_{1}(s)\\
    & = \frac{q^3B}{(2\pi)} \frac{\tau}{\omega_c}  \sum_{n=-\infty}^{\infty} \frac{v^{\mu}_nv^{\nu}_{-n}}{1+in \omega_c  \tau}E^{\nu},
\end{split}
\end{equation}
where $(\mu,\nu)=x,y$, and we have defined the cyclotron frequency as:
\begin{equation}\label{eq:cyclotron_frequency}
    \omega_c \equiv 2\pi/s_0.
\end{equation}
We have also made use of the periodicity of velocity under $s\rightarrow s+s_0$ to write it in terms of Fourier series, $\bv(s)=\sum_{n=-\infty}^{\infty}\bv_ne^{-in \omega_c s }$.

To show that the second line of Eq.~(\ref{eq:current}) holds, note that at every $\bk$ we can define a local coordinate system $(\hat{e}_{v},\hat{e}_{s})$ such that $\bv \equiv v\hat{e}_{\bv}$ where $v\ge 0$, and $\hat{e}_{s}=\hat{e}_{\bv}\times\hat{z}$. The infinitesimal wavevector can be equivalently written as:
\[
\mathrm{d}\bk=\mathrm{d}k_x \hat{e}_x + \mathrm{d}k_y \hat{e}_y= \mathrm{d}k_s \hat{e}_s + \mathrm{d}k_v \hat{e}_v.
\]
Eq.~(\ref{eq:trajectory}) can then be written as ${\mathrm{d}\bk}/{\mathrm{d} s} = qvB \hat{e}_s$, or equivalently $\mathrm{d} k_s = qvB\mathrm{d}s$. As a result,
\[
    \int \mathrm{d} k_x \mathrm{d}k_y = \int \mathrm{d} k_s \mathrm{d}k_v = qB \int  \mathrm{d}\varepsilon \mathrm{d} s.
\]

The conductivity tensor is therefore given by the following expression:
\begin{equation} \label{eq:boltzmann_conductivity}
    \sigma^{\mu\nu} = \frac{q^3B}{2\pi} \frac{\tau}{\omega_c} \sum_{n=-\infty}^{\infty} \frac{v^{(\mu)}_nv^{(\nu)}_{-n}}{1+in \omega_c\tau}.
\end{equation}
Eq.~(\ref{eq:boltzmann_conductivity}) gives the magnetoconductivity for a given FS contour. In the case of multiple FS contours and multiple bands --as due to spin and valley degeneracy in TBG-- conductivities from different FS contours and bands add. Finally, the MR tensor is obtained by inverting the conductivity tensor, i.e., $\rho = \left(\sum_{n,i}\sigma_{n,i}\right)^{-1}$, where $n,i$ are band and contour labels respectively for a given energy level. 

To better understand Eq.~(\ref{eq:boltzmann_conductivity}) consider an example of a parabolic dipsersion with $\varepsilon_{\bk}=\frac{1}{2m_0}(k_x^2+k_y^2)$, where $m_0$ is the bare electron mass. At a fixed energy  $\mu$ the contour is a circle parameterized as: $(k_x,k_y)=\sqrt{2m_0\mu}(\cos\theta,\sin\theta),\ \theta\in[0,2\pi)$. Using method of characteristics, we get: $\frac{\mathrm{d}\theta}{\mathrm{d} s} = -\frac{qB}{m_0}$, or $\theta = \theta_0 - \omega_0 s$, where $\omega_0\equiv \frac{qB}{m_0}$ is the cyclotron frequency of bare electrons. This leads to the periodicity in $s$ to be $s_0=2\pi/\omega_0$, where we have chosen the clockwise trajectory such that $s_0>0$. The Fourier series of the velocity along the constant energy contour is given by: $v_{x}(s)=\sqrt{\frac{\mu}{2m}}\left(e^{-i \omega_0 s}+e^{i\omega_0 s}\right)$, and $v_{y}(s)=\sqrt{\frac{\mu}{2m}}\frac{1}{i}\left(e^{-i \omega_0 s}-e^{i\omega_0 s}\right)$. Substituting into Eq.~(\ref{eq:boltzmann_conductivity}), we obtain the conductivity tensor:
\begin{equation}
    \sigma = q^2\tau \frac{\mu}{2\pi} \frac{1}{1+\omega_0^2\tau^2}
    \begin{pmatrix} 1 & -\omega_0\tau\\ \omega_0\tau & 1  \end{pmatrix}.
\end{equation}
Note that the total number density of filled electrons is given by
    $n = \int \frac{\mathrm{d}^2\bk}{(2\pi)^2} \Theta(\mu-\varepsilon_{\bk}) = \frac{m_0\mu}{2\pi}$.
We therefore reproduce the well known magnetoconductivity tensor:
\begin{equation}
    \sigma = \frac{nq^2\tau}{m_0} \frac{1}{1+\omega_0^2\tau^2}
    \begin{pmatrix} 1 & -\omega_0\tau\\ \omega_0\tau & 1  \end{pmatrix}. 
\end{equation}

In this simple example of a closed FS, the longitudinal resistivity is given by $\frac{m_0}{nq^2\tau}$, independent of the magnetic field. The average of the velocity field, $\bv_{n=0}\equiv \frac{1}{s_0}\int_0^{s_0}\mathrm{d}s \bv(s)$, vanishes. However, for an open FS generally $\bv_{n=0} \neq \mathbf{0}$, i.e.. electrons have a finite drift velocity when traversing the contour due to a magnetic field (see SM Fig.~2). The impact of such a finite drift velocity on the magnetotransport can be qualitatively understood using the following example: in the expression for the conductivity tensor (Eq.~(\ref{eq:boltzmann_conductivity})), we consider $v^{x}_{n=0}\neq 0$ but $v^{y}_{n=0}= 0$. This corresponds to an open FS with a drift velocity along the $x$ direction. In the high field limit ( $\omega_c\tau\propto B \gg 1)$, we truncate the Fourier series at the leading order, and as a result, 
\begin{equation}
    \sigma_{\text{open FS}} \approx \frac{q^3B}{2\pi}\frac{\tau}{\omega_c}\begin{pmatrix}{(v^{x}_0)^2} & -\frac{2\text{Im}(v^{x}_{-1}v^{y}_{1})}{\omega_c\tau}\\ \frac{2\text{Im}(v^{x}_{-1}v^{y}_{1})}{\omega_c\tau} & \frac{|v^{y}_1|^2}{\omega_c^2\tau^2}\end{pmatrix},
\end{equation}
where we made use of the equality: $\bv_{-n}=\bv_{n}^*$. Inverting the matrix, we obtain the MR tensor:
\begin{equation}
\begin{split}
    {\rho}_{\text{open FS}} &\approx \frac{(2\pi)\omega_c}{q^3B \tau} \frac{1}{4\text{Im}(v_{-1}^{x}v_{1}^{y})^2+(v_0^{x})^2|v_{1}^{y}|^2} \\
    & \times \begin{pmatrix}|v^{y}_1|^2 & {2\text{Im}(v^{x}_{-1}v^{y}_{1})}{\omega_c\tau}\\ -{2\text{Im}(v^{x}_{-1}v^{y}_{1})}{\omega_c\tau} & (v^{(x)}_0)^2\left(\omega_c\tau\right)^2\end{pmatrix}.
\end{split}
\end{equation}
It is clear that $\rho_{yy}\propto B^2$ whereas $\rho_{xx}\sim \mathcal{O}(1)$. We therefore arrive at the important conclusion that for an open FS, the longitudinal MR has non-saturating $B^2$ behavior along the axis with a zero drift velocity ($\hat{y}$ in the above example), and saturating behavior along the other axis. 

\subsection{Magnetotransport in TBG under heterostrain}

\begin{figure*}
    \centering
    \includegraphics[width=0.8\linewidth]{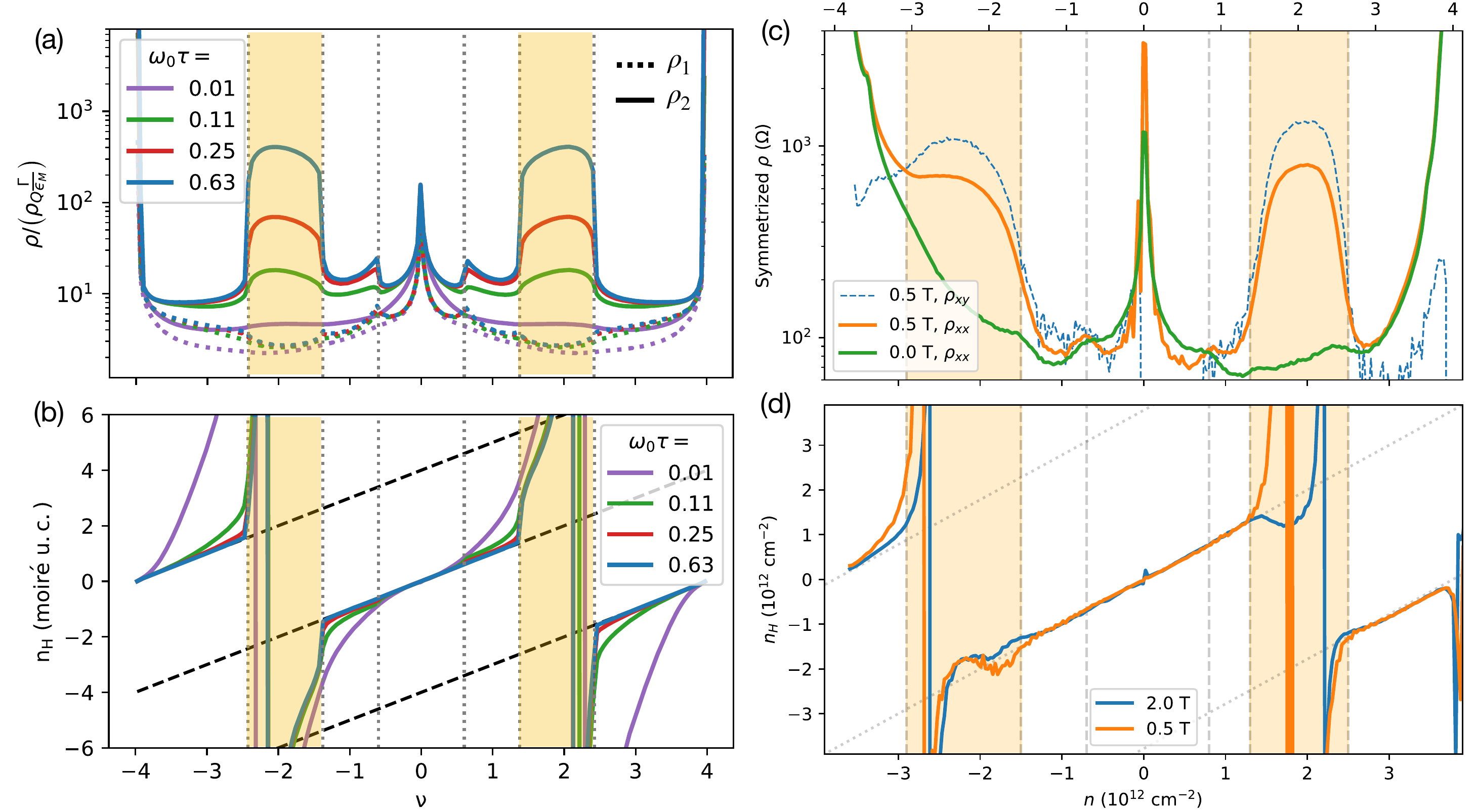}
    \caption{Magnetotransport properties of strained TBG. {\bf(a-b)} Theoretical calculations of transport properties as a function of magnetic field strengths $\omega_0\tau$ for $\theta=1.38^\circ$, $\epsilon=0.2\%$ and $\varphi=0^\circ$. The cyclotron frequency $\omega_c$ defined in Eq.~(\ref{eq:cyclotron_frequency}) is filling-dependent, hence our choice to use the bare cyclotron frequency $\omega_0\tau=eB\tau/m_0$. The vertical dashed lines mark the calculated van Hove points, with yellow regions indicating open FSs. (a) Longitudinal MR along the principal axes $\hat{e}_1$ (dashed) and $\hat{e}_2$ (solid) in units of $\rho_Q\frac{\Gamma}{\epsilon_M}$, where $\rho_Q\equiv h/e^2$ is the quantum of resistance, $\Gamma\equiv\hbar/\tau$ is the transport decay rate, and $\epsilon_M\equiv \hbar v_F|\bK|{\theta}$ is the characteristic energy scale for moir\'e electrons. For a transport rate $\Gamma=0.1\mathrm{meV}$, $\rho_Q\frac{\Gamma}{\epsilon_M}\approx 9.6\Omega$, and $\omega_0\tau\approx 0.13$ is equivalent to a magnetic field strength $B\approx 0.11T$. (b) Hall number $n_H\equiv e\rho_H/B$. {\bf(c-d)} Experimental measurements of longitudinal MR (contact pair 14-15) and transverse MR (contact pair 15 - 5) for the TBG sample in Ref.~\cite{FinneyPNAS2022} at 1.6 K. Vertical dashed lines mark the densities that we ascribe to van Hove points based on the cusp near $\nu\sim 0.8$ and the onset of quadratic MR (shaded yellow). Finite-field resistivities in panel (c) are symmetrized: $\rho=(\rho(B) + \rho(-B)) / 2$. Panel (d) is calculated from the antisymmetrized transverse resistivity.}
    \label{fig:main_fig}
\end{figure*}

We proceed to apply the above results to analyze the magnetotransport in TBG. The theory satisfactorily explains the weak-field magnetotransport measurements presented in Ref.~\cite{FinneyPNAS2022}. We then present two predictions of the theory that we did not anticipate prior to starting this work: the dependence of the principle axis of transport on filling, and the behavior of magnetoresistance and quantum oscillations at densities between the CNP and the onset of quadratic MR. The former is of academic interest, however it cannot be confirmed with our present data sets because of limitations of the Hall bar geometry. The latter can be considered smoking gun evidence for the the presence of the lowest-energy van Hove point and the energetic splitting of the Dirac cones.

We do not expect our strained BM model in Eq.~(\ref{eq:strained_bm}) to yield precise agreement with experiment, so we do not perform fine-tuning of its input parameters. Specifically, the model has particle-hole symmetry, which is absent from experimental measurements. More sophisticated non-interacting model calculations \cite{Kang2022b,Vafek2022a} as well as interaction renormalizations \cite{Choi2021} are likely necessary to properly account for such details. Although the general phenomena of open FSs and quadratic MR holds for a broad range of heterostrain parameters, we present calculations for $\theta=1.38^\circ$, $\epsilon=0.2\%$ and $\varphi=0^\circ$, parameters chosen to yield reasonable quantitative agreement between the theoretical and experimental results both on the filling range of open FSs, as well as on the frequencies of magnetoresistance oscillations to be presented later.

In Fig.~\ref{fig:main_fig}, we show the computed MR along the principal transport axes (a) and the Hall number (b). For comparison, we plot the experimentally measured longitudinal and transverse resistivities (c) and Hall number (d) for the TBG device studied in Ref.~\cite{FinneyPNAS2022}.

In the filling ranges with open FSs, the calculated $\rho_2(B)$ exhibits quadratic non-saturating MR, whereas $\rho_1(B)$ saturates. The filling range for which quadratic MR occurs is bounded by the two outermost van Hove points of the zero-field strained band structure. In experiment, we observe quadratic MR in longitudinal resistivity within a similar range of fillings. More strikingly, we observe quadratic MR in the transverse resistivity as well. In some cases, the symmetric part of the transverse resistivity becomes larger than that of the longitudinal resistivity with field. As discussed earlier, this degree of mixing can be attributed to the misalignment between the strain-induced principal axis of transport and the direction of current flow in the Hall bar geometry.

At the first van Hove point ($\nu\approx \pm 0.6$), the non-analyticity in the density of states leads to a cusp in the first derivative of the zero-field resistivity with respect to filling (see SM Fig.~6). As shown in Fig.~\ref{fig:main_fig}(a), at $B\neq 0$ the longitudinal resistance develops a cusp as a function of filling at the first van Hove point. The cusp becomes more pronounced with increasing $B$. Experimentally as shown in Fig.~\ref{fig:main_fig}(c), there is a cusp-like feature developing at $|\nu|\sim 0.5-0.8$ depending on the contact pair within the device used, consistent with theoretical predictions. In many contact pairs, this feature presents as a shoulder at $B=0$, only developing into a cusp at $B \sim 0.1$ T (see SM Fig.~7).

As depicted in Fig.~\ref{fig:main_fig}(b), the calculated filling dependence of the Hall number shows two singular sign changes inside the open FS regions near $\nu\approx \pm 2$. The sign changing singularity in the open FS region is $B$-independent, and is not directly associated with any van Hove point (see SM Fig.~5 for a plot of $\rho_H(B)$, which crosses zero at the same filling fraction inside the open FS filling range for varying field strength). Moreover, the filling dependence of the Hall number $n_H$ tracks the filling fraction in a broad filling range near the CNP, with the filling range being extended upon increasing $B$. In Fig.~\ref{fig:main_fig}(d), we observe the same general shape of the Hall number. Within the open FS filling range, however, the measured Hall number qualitatively deviates from the theoretical curves. We attribute this to a small constant offset in the magnetic field of order ~10-20 mT, likely resulting from trapped flux in the superconducting magnet. Here a large quadratic symmetric component of the transverse resistivity is concurrent with a vanishing antisymmetric component. An offset of only a few mT will lead to a small part of the symmetric component mixing into the antisymmetric component, leading to these deviations from theory (See SM Fig.~8).

\begin{figure}
    \centering
    \includegraphics[width=\linewidth]{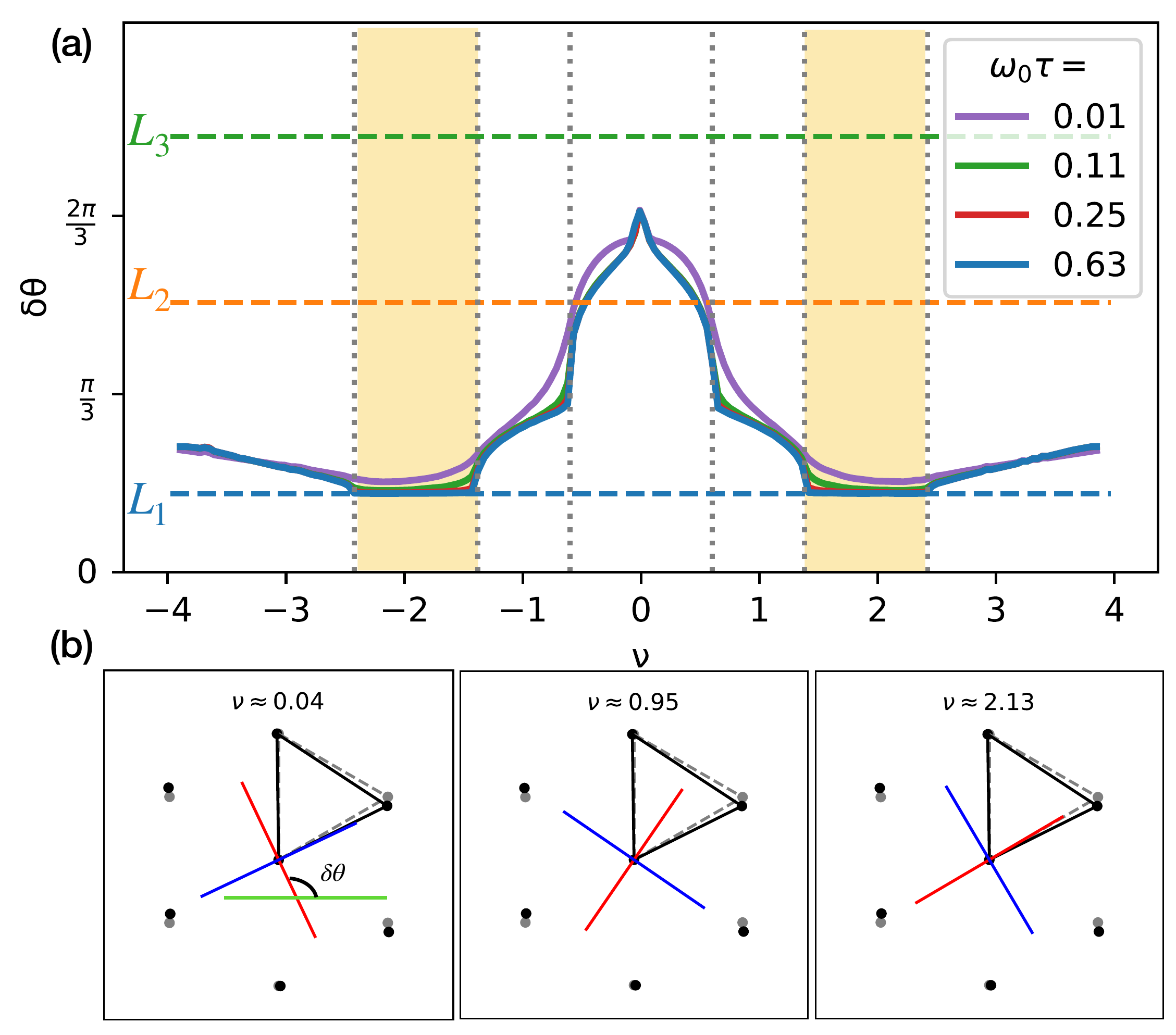}
    \caption{{\bf(a)} Rotation of the transport principal axis $\hat{e}_1$ with respect to the global coordinate system for strained BM with $\epsilon=0.2\%$ and $\varphi=0^\circ$. The three horizontal dashed lines are the bond directions. In the open FS region, the saturating MR axis is locked to the shortest bond  ($\bL_1$) direction. However, it rapidly rotates in the closed FS region upon approaching the CNP. {\bf (b)} Principal transport axes $\hat{e}_1$ (red) and $\hat{e}_2$ (blue) for a few filling fractions. Near the CNP, $\hat{e}_1$ is perpendicular to the shortest moir\'e bond direction. In the open FS filling range (e.g. $\nu\approx 2.13$) it is rotated to be parallel to the shortest bond direction. }
    \label{fig:strained_principal_rotation}
\end{figure}

Our calculation finds a dramatic rotation of the principal axis with filling, as illustrated in Fig.~\ref{fig:strained_principal_rotation}. In the filling range with open FSs, the principal axis with saturating MR ($\hat{e}_1$) is aligned with direction of the shortest moir\'e triangular bond, suggesting that the electrons are hopping more efficiently along the shortest bond, which leads to a larger conductivity and therefore a smaller resistivity. Interestingly, when filling is changed from the second van Hove point ($\nu\approx\pm 1.3$) to the vicinity of the CNP, $\hat{e}_1$ rotates dramatically to the perpendicular direction compared to the filling range with open FSs. The rotation of the principal axis is likely due to the opposite elongation of the larger Fermi pocket encircling a Dirac point compared to the open FS contours, see for example Figs.~\ref{fig:strain_energetics}(d-f). The rotation of the transport axis with filling purely due to strain-induced bandstructure effects demonstrates that filling dependence of the principal axes orientation need not be associated with interaction induced nematicity \cite{Cao2020nematicity}. Such a filling-dependent rotation of the principal transport axis was not possible to observe in Ref.~\cite{FinneyPNAS2022} using the Hall bar geometry, where only $\rho_{xx}$ and $\rho_{yx}$ are measured but not $\rho_{yy}$. Additional transport measurements are needed, where the filling-dependence of the entire resistivity tensor can be mapped out.

\begin{figure*}
    \centering
    \includegraphics[width=0.7\linewidth]{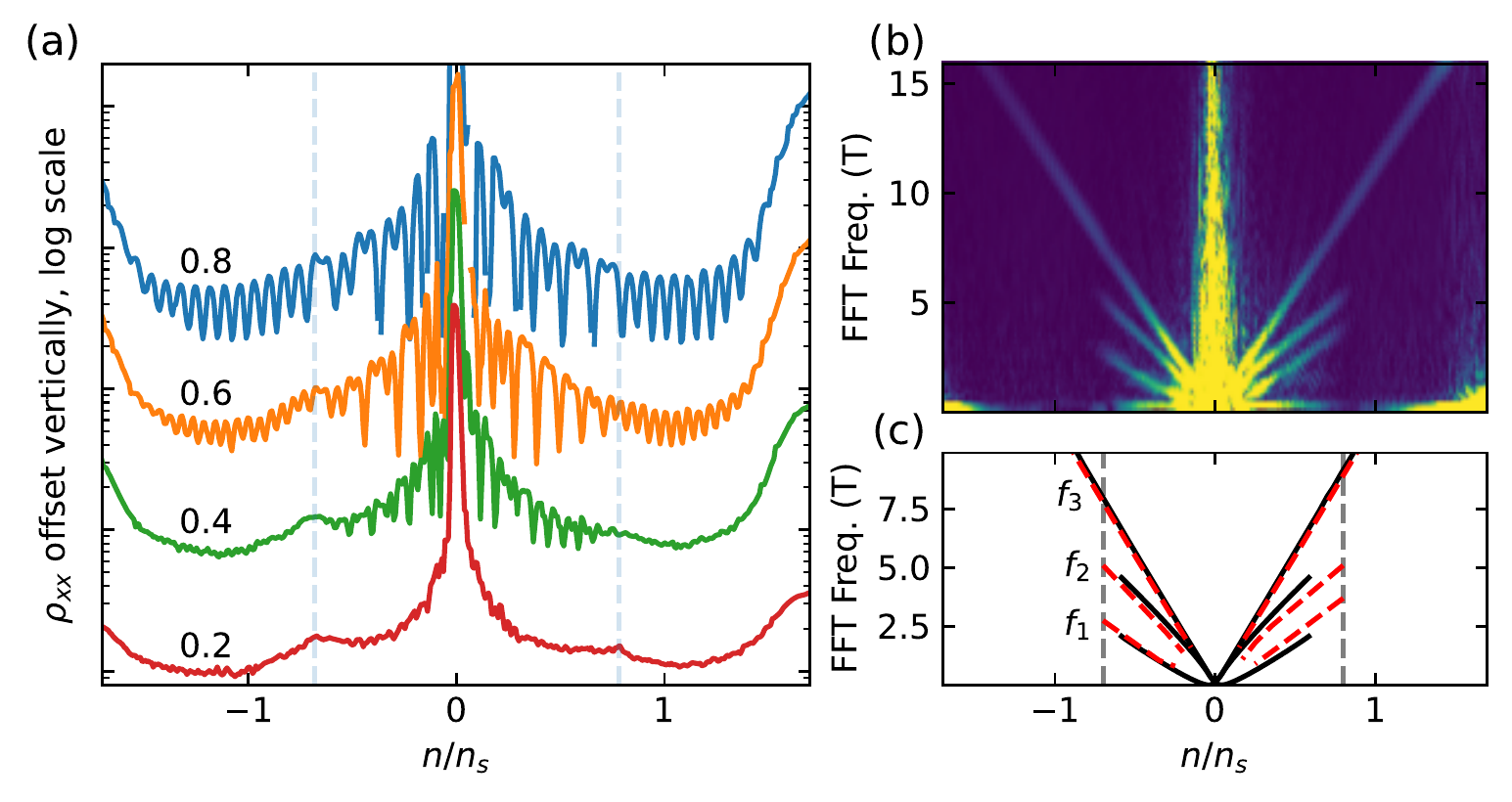}
    \caption{{\bf(a)} Line cuts of MR near the CNP taken at 26 mK in contact pair 4 - 5 at the indicated field strengths, in Tesla. Vertical dashed lines indicate our estimated location of the lowest-energy van Hove points, based on the cusps in resistivity at low field. Within the region bounded by these points, the quantum oscillations show up before 0.4 T, and their relative strengths do not follow a simple pattern. Outside of this region, the quantum oscillations onset at higher field, and every multiple of 4 quantum Hall filling fraction is observed relatively equally. {\bf(b)} Fourier transform of the quantum oscillation data with respect to $1/B$. It reveals a transition from two pockets to one pocket at the lowest-energy van Hove points. {\bf (c)} Schematic description of the frequencies observed in panel (b). Red dashed lines are frequencies from the experimental data. Solid black lines are predictions from the theory for $\epsilon=0.2\%$ and $\varphi=0^\circ$. The two frequencies $f_1$ and $f_2$ sum to the one-pocket frequency $f_3$ that extends beyond the first van Hove point. They additionally account for the nontrivial relative strengths of the quantum oscillations within the bounds of the first van Hove points. As with other details of this work, the theory predicts electron-hole symmetry, while some asymmetry is observed in experiment.}
    \label{fig:onsetQO}
\end{figure*}

Since this theory predicts a third van Hove point between the CNP and the filling range with open FSs, a direct measurement of this van Hove point is desired. In Fig.~\ref{fig:onsetQO} we reanalyze quantum oscillation measurements of the TBG device discussed in Ref.~\cite{FinneyPNAS2022}. The effective cyclotron mass $m^*$ is light in the filling range with two small closed Fermi pockets, and dramatically heavier in the filling range with only one closed pocket (Figs.~\ref{fig:strain_energetics}(d-f) and SM Fig.~5). The large difference in masses on either side of the innermost van Hove singularity can be used to explain the substantially earlier onset of quantum oscillations with increasing field close to the CNP than away from it, as shown in Fig.~\ref{fig:onsetQO}(a). Fig.~\ref{fig:onsetQO}(b) is a Fourier transform of the quantum oscillation data with respect to $1/B$. In the filling range of $-0.7\le \nu\le 0.8$ three distinct frequencies $f_{i=1,2,3}$ are clearly observed in the data, with $f_1$ and $f_2$ corresponding to two small Fermi pockets, and $f_3=f_1+f_2$ to the breakdown orbit when the inverse magnetic length is comparable to the momentum space distance between the two small Fermi pockets \cite{Schoenberg2009}. Outside of the filling range only $f_3$ is observed, showing that there are Lifshitz transitions, one on either side of the CNP, that we ascribe to crossing the lowest-energy van Hove points. Furthermore, these filling fractions also correspond to the cusp-like features in the longitudinal MR data shown in Fig.~\ref{fig:main_fig}(c), consistent with theoretical predictions for its behavior at van Hove singularities. Therefore, the quantum oscillation data unambiguously demonstrates the existence of a third van Hove singularity at filling fractions between the CNP and the filling range of $B^2$ MR. It is interesting to note again, that the Hall number does not show a sign-changing singularity at this van Hove point, as illustrated in Fig.~\ref{fig:main_fig}(b) and (d).

The frequencies $f_{1,2}$ are a strong constraint on the amount of heterostrain in the TBG sample. Specifically, as illustrated in Fig.~\ref{fig:onsetQO}(c), the frequency $f_2$ is roughly two times $f_1$, showing that the two small Fermi pockets have an area ratio $\sim 2:1$. Theoretically as illustrated by the solid black lines in Fig.~\ref{fig:onsetQO}(c), for a heterostrain strength $\epsilon=0.2\%$ and $\varphi=0^\circ$, the areas $A_{i=1,2}$ of the two small pockets, when converted to frequency via $f_i^{-1}\equiv(\Delta\frac{1}{B})_i=\frac{2\pi e}{\hbar A_i}$, are in good agreement with experiment.

We observe behavior qualitatively similar in all respects to that in Fig.~\ref{fig:onsetQO}(a) in all 3 longitudinal contact pairs for which we have dilution-fridge measurements (see SM Fig.~11).

In addition to the quantum oscillation measurements above, we propose an additional experimental procedure for identifying the van Hove points. As usual, at van Hove singularities there are non-analyticities in the electronic density of states. Such non-analyticities will lead to cusps in the first derivative of the zero-field resistivity with respect to filling (see SM Fig.~6). This can be probed via transport measurements, for example, by adding a small ac modulation of the filling or by numerical differentiation of the dc data.

\section{Summary and Outlook}
In summary, we have shown that due to the large size of the moir\'e unit cell at small twist angles, even a small amount of uniaxial heterostrain on the microscopic scale can lead to dramatic changes in the narrow bands of twisted bilayer graphene. A key feature of the strained bandstructure is the splitting of the respective energetic degeneracies of the two Dirac points and the three van Hove points. The splitting of the two Dirac points leads to a semimetallic state with two small Fermi pockets at the CNP. On the other hand, the two outermost van Hove points bound a broad filling range near $\nu=\pm 2$ where the constant energy contours become open. Interestingly, the elongation of the larger Fermi pocket near the CNP is perpendicular to that of the open FSs, the latter being perpendicular to the direction of the shortest moir\'e triangular bond.

We have analyzed the resulting magnetotransport in strained TBG in the framework of the Boltzmann equation using the method of characteristics, treating the magnetic field non-perturbatively. We showed that a non-saturating quadratic longitudinal magnetoresistance in a broad filling range near $\nu=\pm 2$ naturally arises due to the heterostrain-induced open Fermi surfaces, therefore explaining the experimental results in the off-magic-angle devices \cite{FinneyPNAS2022}. We have also shown that the sign-changing singularities in the Hall number occur in the open FS filling range and are not directly associated with any van Hove singularity as commonly assumed, e.g., in Ref.~\cite{Park2021}. Furthermore, our results reveal a dramatic rotation of the transport principal axis as the filling is tuned from the charge neutrality point to the filling range of open Fermi surfaces. This is entirely attributed to the strained non-interacting bandstructure effects, and does not require interaction-induced electronic nematicity for explanation.

Given the importance of energy-shifted van Hove points in the transport properties of TBG devices, we have analyzed previous quantum oscillation data, which has revealed a Lifshitz transition from two pockets to one pocket at a filling fraction where the innermost van Hove singularity is predicted to occur based on theoretical calculations, therefore offering strong evidence of heterostrain effects on these devices. We have further proposed several additional signatures to look for in future experiments. These include cusps in the derivative of zero field resistivity with respect to filling, a significant difference in cyclotron mass on either side of the innermost van Hove singularity, and a principal transport axis with saturating magnetoresistance in the open Fermi surface filling range. 

Finally, given the amplifying effect of a small strain at the underlying carbon lattice scale on the moir\'e lattice scale, the latter of which controls the electronic behavior within the narrow bands, it is tantalizing to consider strain engineering of such devices to achieve effects which would be impossible in regular solids due to structural instabilities.

\section{Acknowledgement}
 \textbf{Funding}: X.W. acknowledges financial support from National MagLab through Dirac fellowship, which is funded by the National Science Foundation (Grant No. DMR-1644779) and the state of Florida. O.V. was supported by NSF Grant No.~DMR-1916958 and is partially funded by the Gordon and Betty Moore Foundation's EPiQS Initiative Grant GBMF11070, National High Magnetic Field Laboratory through NSF Grant No.~DMR-1157490 and the State of Florida. Device measurements and analysis were supported by the U.S. Department of Energy, Office of Science, Basic Energy Sciences, Materials Sciences and Engineering Division, under contract DE-AC02-76SF00515. Measurement infrastructure was funded in part by the Gordon and Betty Moore Foundation’s EPiQS Initiative through grant GBMF3429 and grant GBMF9460. D.G.-G. gratefully acknowledges support from the Ross M. Brown Family Foundation. Sandia National Laboratories is a multimission laboratory managed and operated by National Technology \& Engineering Solutions of Sandia, LLC, a wholly owned subsidiary of Honeywell International Inc., for the U.S. Department of Energy’s National Nuclear Security Administration under contract DE-NA0003525. K.W. and T.T. acknowledge support from JSPS KAKENHI (Grant Numbers 19H05790, 20H00354 and 21H05233). Part of this work was performed at the Stanford Nano Shared Facilities (SNSF), supported by the National Science Foundation under award ECCS-2026822. 

\textbf{Author contributions}: O.V. conceived the theoretical explanation of the experiments. X.W. and O.V. performed calculations. J.F., L. R., and C.H. fabricated devices. J.F. and X.W. analysed the data. K.W. and T.T. generously supplied the hBN crystals. A.L.S., M.A.K., O.V., and D.G.-G. supervised the experiments and analysis. The manuscript was prepared by X.W. and J.F. with input from all authors.

\bibliographystyle{apsrev4-2}
\bibliography{openFS}


\pagebreak

\onecolumngrid
\begin{center}
  \textbf{\large Supplementary Materials for ``Unusual magnetoresistance in twisted bilayer graphene from strain induced open Fermi surfaces"}
\end{center}

\setcounter{section}{0}
\setcounter{equation}{0}
\setcounter{figure}{0}
\setcounter{table}{0}
\setcounter{page}{1}

We present additional theoretical results and experimental measurements in support of the main text.
\section{Heterostrain effects on the geometry of moir\'e superlattice}
Our off-magic-angle twisted bilayer graphene (TBG) devices in Ref.~\cite{FinneyPNAS2022} are prepared using the ``tear-and-stack" procedure, and as a result, strain is inevitably introduced. Here we first show that while the moir\'e unit cell vectors are strongly deformed by even an infinitesimal amount of uniaxial heterostrain in the device, the unit cell area is much less affected. As a result, for the off-magic-angle device studied in Ref.~\cite{FinneyPNAS2022}, we can have a good estimate of the twist angle ($\theta$) based on the moir\'e unit cell area alone.

In the limit of small deformations, both the uniaxial heterostrain and a small twist angle are captured via a coordinate transformation: $\br'_l = \br + \bu_l(\br)$, where $l=t,b$ labels the top (bottom) graphene layers, and $\bu_l(\br) \approx \mathcal{E}_l \br$ is the local deformation field. The symmetric and antisymmetric part of the $2\times 2$ tensor $\mathcal{E}_l$ describes strain and rotation respectively. For twist angle $(\theta)$ and a uniaxial heterostrain of strength $(\epsilon)$ and direction $(\varphi)$, we parameterize $\mathcal{E}_{t}=-\mathcal{E}_{b}\equiv \mathcal{E}/2$, where $\mathcal{E}\equiv \mathcal{T}(\theta)+ \mathcal{S}(\epsilon,\varphi)$, and given by: 
\begin{equation}
\mathcal{T}(\theta) = \begin{pmatrix} 0 & -\theta\\ \theta & 0\end{pmatrix},\ \mathcal{S}(\epsilon,\varphi) = R_{\varphi}^T \begin{pmatrix} -\epsilon & 0\\ 0 & \nu\epsilon \end{pmatrix}R_{\varphi}.
\end{equation}
Here $R_\varphi$ is the two-dimensional rotation matrix, and $\nu\approx 0.16$ is the Poisson ratio  \cite{Kerelsky2019}. Physically, $\epsilon>0$ corresponds to compressing the top layer while streching the bottom layer along the $x$-axis. A relative deformation $\mathcal{E}$ between the graphene bilayers generate a moir\'e superlattice, with moir\'e reciprocal lattice vectors given by:
\begin{equation}\label{eq:reciprocal_latt_vectors}
    \bg_{i=1,2}=\mathcal{E}^T\bG_{i=1,2},
\end{equation}
where $\bG_{i}$ are reciprocal lattice vectors of the undeformed monolayer graphene. Eq.~(\ref{eq:reciprocal_latt_vectors}) can be used to uniquely determine the three parameters $(\theta,\epsilon,\varphi)$. Additionally it also determines a global angle $\alpha$ that measures the rotation between the lab and theoretical coordinate systems.

Uniaxial heterostrain has a dramatic effect on the distortion of the moir\'e unit cell vectors, as $|\delta \bg|/|\bg|\sim \mathcal{O}(\epsilon/\theta)$. However, its effect on the moir\'e unit cell area is much smaller. To show this, note that the area of the moir\'e Brillouin zone is calculated as: 
\begin{equation}
    A_{mBZ} = \left| (\bg_1 \times \bg_2 )\cdot \hat{z}\right| = \left| \bg_1^T (i\sigma_y) \bg_2 \right|,
\end{equation}
where on the second equality we have used a vector notation $\bg_i\equiv(g_{i,x},g_{i,y})^T$. Following Eq.~(\ref{eq:reciprocal_latt_vectors}), we obtain that the area of the moir\'e Brillouin zone is independent on $\varphi$, and calculated as:
\begin{equation}
    A_{mBZ} = (\theta^2-\nu^2\epsilon^2)A_{BZ},
\end{equation}
where $A_{BZ}\equiv \left|(\bG_1\times\bG_2)\cdot\hat{z}\right|$ is the Brillouin zone area of the undeformed monolayer graphene. The area of the strained moir\'e unit cell can be calculated in a similar manner, and we get: $A_{m.u.c.}=A_{u.c.}/(\theta^2-\nu^2\epsilon^2)$, where $A_{u.c.}$ is the unit cell area of undeformed monolayer graphene. Observe that the heterostrain only affects the area of the moir\'e unit cell by $\mathcal{O}(\nu^2\epsilon^2/\theta^2)$ which is much smaller than the linear distortion of moir\'e unit cell vectors. 

With only a knowledge of the moir\'e unit cell areas in Ref.~\cite{FinneyPNAS2022} (see Table~\ref{tab:tab_exp}), we estimate the twist angle to be $\theta \sim 1.35^\circ - 1.39^\circ$ for various contact pairs studied using the Hall bar geometry.

\section{Constraining heterostrain from transport measurements}
\begin{table}[h]
    \centering
    \begin{tabular}{c|c|c|c|c|c|c|c}
contact pairs & unit cell area (nm$^2$) & $\nu_1$ & $\nu_2$ & $\nu_3$ & $\nu_4$ & $\nu_5$ & $\nu_6$\\
& $\pm 0.1$ & $\pm 0.05$ & $\pm 0.05$ & $\pm 0.05$ & $\pm 0.05$ & $\pm 0.05$ & $\pm 0.05$\\
\hline 
4 - 5 & 95.0 & -2.95 & -1.58 & -0.74 & 0.84 & 1.47 & 2.42\\
\hline
5 - 6 & 91.5 & -3.28 & -1.75 & -0.66 & 0.76 & 1.53 & 2.84\\
\hline
6 - 7 & 89.2 & -3.70 & -1.68 & -0.39 & 0.50 & 1.57 & 2.92\\
\hline
7 - 8 & 91.8 & -3.27 & -1.63 & -0.49 & 0.60 & 1.42 & 2.94\\
\hline
14 - 15 & 93.6 & -3.10 & -1.60 & -0.75 & 0.85 & 1.50 & 2.46\\
\hline
15 - 16 & 90.5 & -3.32 & -1.66 & -0.66 & 0.83 & 1.66 & 2.65\\
\hline
16 - 17 & 90.1 & -3.33 & -1.89 & -0.44 & 0.50 & 1.78 & 2.89\\
\hline
17 - 18 & 91.8 & -3.38 & -1.63 & -0.49 & 0.54 & 1.53 & 2.94

\end{tabular}
    \caption{Table of moir\'e unit cell areas and filling fractions $\nu_{i=1\dots6}$ of the six van Hove singularities for different contact pairs of the Hall bar measurements in Ref.~\cite{FinneyPNAS2022}. The filling fractions are obtained by-eye based on magnetotransport measurements (Fig.~\ref{fig:allxx}). Theoretical calculations predict non-analytic behaviors of longitudinal magnetoresistance at all van Hove singularities.}
    \label{tab:tab_exp}
\end{table}

For the TBG device studied in Ref.~\cite{FinneyPNAS2022}, the deformed moir\'e lattice vectors were not measured. Nevertheless, here we show that magnetotransport measurements, along with theoretical calculations based on the strained Bistrizer-MacDonald (BM) Hamiltonian, offer strong constraints on the heterostrain in the device. We caution, however, that since the strained BM model is an approximate description of the narrow bands of TBG, a precise determination of heterostrain from model calculations is not feasible. 

First of all, as predicted by theoretical calculations, the van Hove singularities of the band structure lead to non-analytic behavior for the longitudinal magnetoresistance as a function of electron filling. The filling fractions for the six van Hove singularities in the narrow band are listed in Table~\ref{tab:tab_exp} for various contact pairs. Secondly, magnetic oscillations show a Lifshitz transition at the inntermost van Hove singularities ($\nu_3,\nu_4$), from two small Fermi pockets closer to the charge neutrality point to one Fermi pocket away from it. Furthermore, the areas of the two small Fermi pockets, as revealed by the frequencies of magnetic oscillations, show a $2:1$ or smaller ratio. Both the filling fractions for van Hove singularities and the pocket area size offer strong constraints for the heterostrain. Qualitatively, on the one hand, a broader filling range of open Fermi surfaces can be achieved by increasing the strength of uniaxial heterostrain. On the other hand, to obtain Fermi pocket area sizes near $2:1$ ratio or smaller, a smaller heterostrain is necessary as it leads to a weaker splitting of the two Dirac cones. For theoretical calculations presented in the main text, we find $\epsilon=0.2\%$ and $\varphi=0^\circ$ to give reasonably good agreements with both experimental observations described above. A larger heterostrain strength ($\epsilon=0.3\%$) will lead to a much larger pocket area ratio ( $4:1$ for $\epsilon=0.3\%$ and $\varphi=0^\circ$), inconsistent with magnetic oscillation measurements. On the other hand, a smaller heterostrain strength $\epsilon=0.1\%$ decreases the filling range of open Fermi surfaces dramatically, inconsistent with the longitudinal magnetoresistance measurements. 

\section{Detailed band structure analysis for varying uniaxial heterostrain}
\begin{figure}
    \centering
    \includegraphics[width=0.7\linewidth]{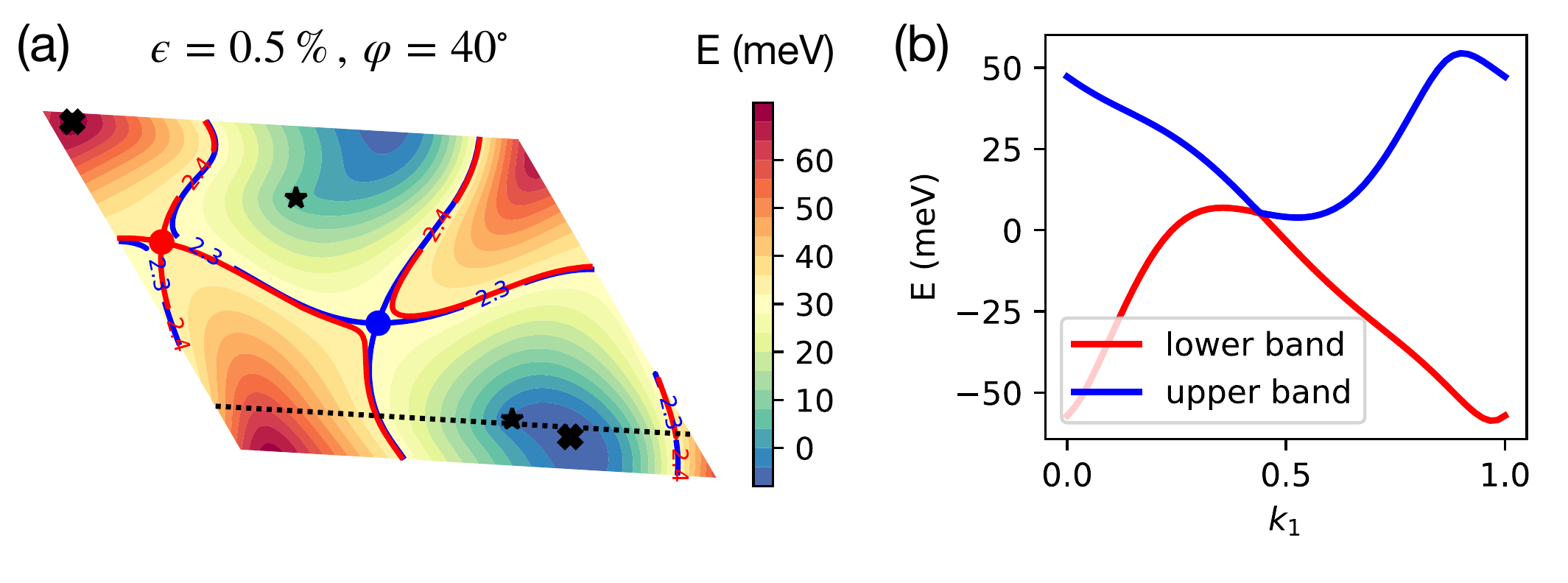}
    \caption{Type II Dirac cone can occur at larger strengths of heterostain. Here we show an example of a type II Dirac cone for $\epsilon=0.5\%$ and $\varphi=40^\circ$. (a) is the energy contour map of the upper band in graphene valley $\bK$, and (b) is the line cut corresponding to the black dotted line in (a).}
    \label{fig:typeIIDP}
\end{figure}

In the main text we discussed the band structure of the strained TBG for $\epsilon=0.2\%$. The main effect of uniaxial heterostrain is to break the respective energetic degeneracies of the two Dirac points and three van Hove points of a given band, therefore giving rise to a semimetallic state at charge neutrality point, and open Fermi surface regions bounded by the two outermost van Hove points. However for a larger heterostrain, the innermost van Hove point moves closer to one of the Dirac point. As a result, both Dirac cones become type II titled, and the innermost van Hove points of both the upper and lower bands are annihilated. In turn two new band extrema are formed. This is illustrated in Fig.~\ref{fig:typeIIDP}.

We also explore the possibilities of heterostrain-induced higher order van Hove singularities which is possible for the magic-angle TBG as discussed in Ref.~\cite{ZhenBi2019}. We checked that for $\theta=1.38^\circ$, and up to uniaxial heterostrain strength of $\epsilon=0.7\%$, no higher order van Hove singularities are found. This shows that the band flattening effect at the magic angle may be important for strain engineering of higher order van Hove points.

\begin{figure}
    \centering
    \includegraphics[width=0.35\linewidth]{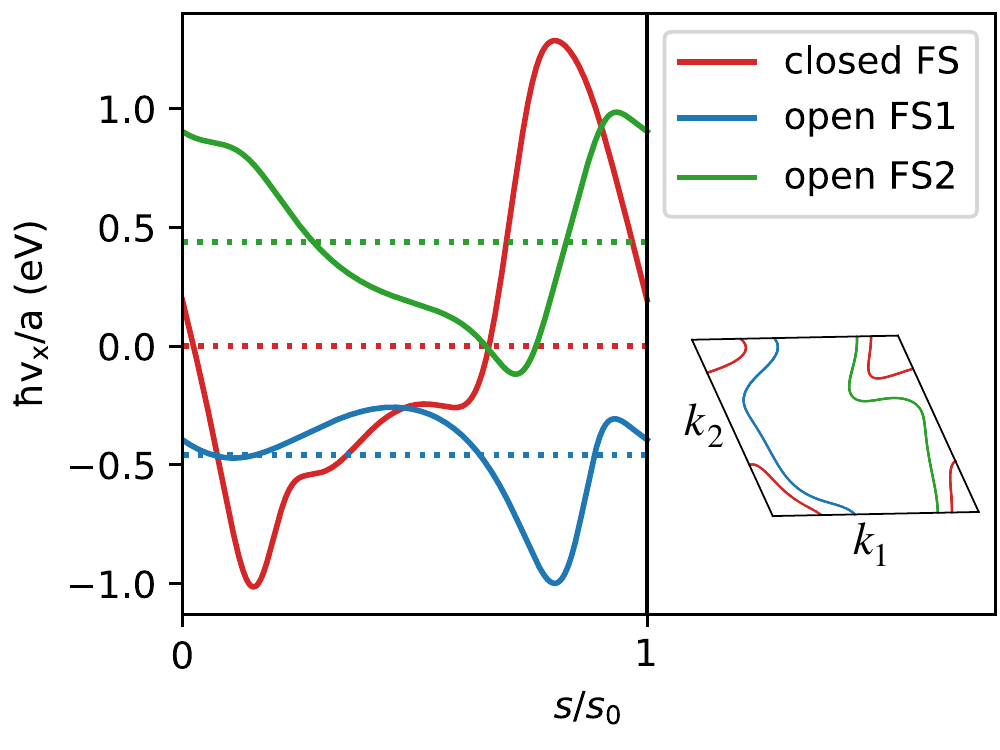}
    \caption{Velocity field of typical closed and open Fermi surfaces. Whereas for a closed Fermi surface the averaged velocity vanishes, for open Fermi surfaces this is generally violated. }
    \label{fig:transport_strained}
\end{figure}

In Fig.~\ref{fig:transport_strained} we plot the velocity fields $v_x(s)$ and $v_y(s)$ on typical open and closed Fermi surfaces for the strained TBG, parameterized by $s\in[0,s_0)$ as defined in the main text. For the closed Fermi surface contours, the averaged velocity, $\bv_{n=0}\equiv \frac{1}{s_0}\int_0^{s_0}\mathrm{d}s \bv(s)$, is zero. On the other hand, for a typical open Fermi surface contour, it is finite, and as a result the electron traversing the open Fermi surface contour in the presence of a magnetic field has a finite drift velocity. As discussed in the main text, this is the reason for the non-saturating $B^2$ magnetoresistivity (MR) observed in strained TBG devices.

\section{More details on magnetotransport in TBG}

\begin{figure}
    \centering
    \includegraphics[width=0.9\linewidth]{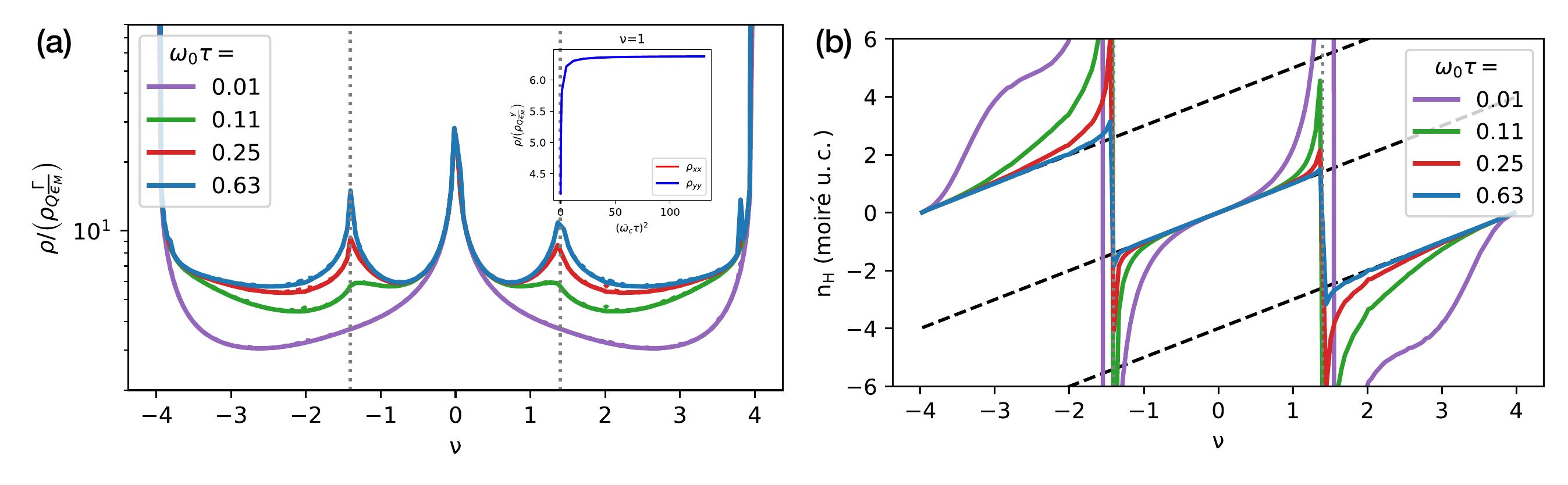}
    \caption{(a) Longitudinal resistivitities $\rho_{xx}$ (dashed) and $\rho_{yy}$ (solid) for unstrained BM model as a function of filling $\nu$. Different colors represent varying magnetic field strength. The inset shows a saturating MR as magnetic field is increased. (b) Hall number as a function of filling fraction. The magnetic field increases from blue to red. Gray dotted vertical lines mark positions of the van Hove singularities.}
    \label{fig:transport_unstrained}
\end{figure}

Here we show that while\ $B^2$ longitudinal MR generally occurs for strained TBG due to open Fermi surfaces, it does not occur for unstrained devices. In Fig.~\ref{fig:transport_unstrained}, the longitudinal MRs $\rho_{xx}$ and $\rho_{yy}$ as well as the Hall number $n_H$ are plotted for an unstrained BM model calculation. First of all, $\rho_{xx}=\rho_{yy}$ since the unstrained TBG has $C_{3z}$ rotational symmetry. Secondly, cusp-like features develop at the triply-degenerate van Hove point at filling fractions $\nu\approx\pm1.4$, and are attributed to the non-analyticities in the density of states behavior at the van Hove singularities. Finally, unstrained TBG has saturating MR across all filling range, as illustrated in the inset to Fig.~\ref{fig:transport_unstrained}(a).

\begin{figure}
    \centering
    \includegraphics[width=0.9\linewidth]{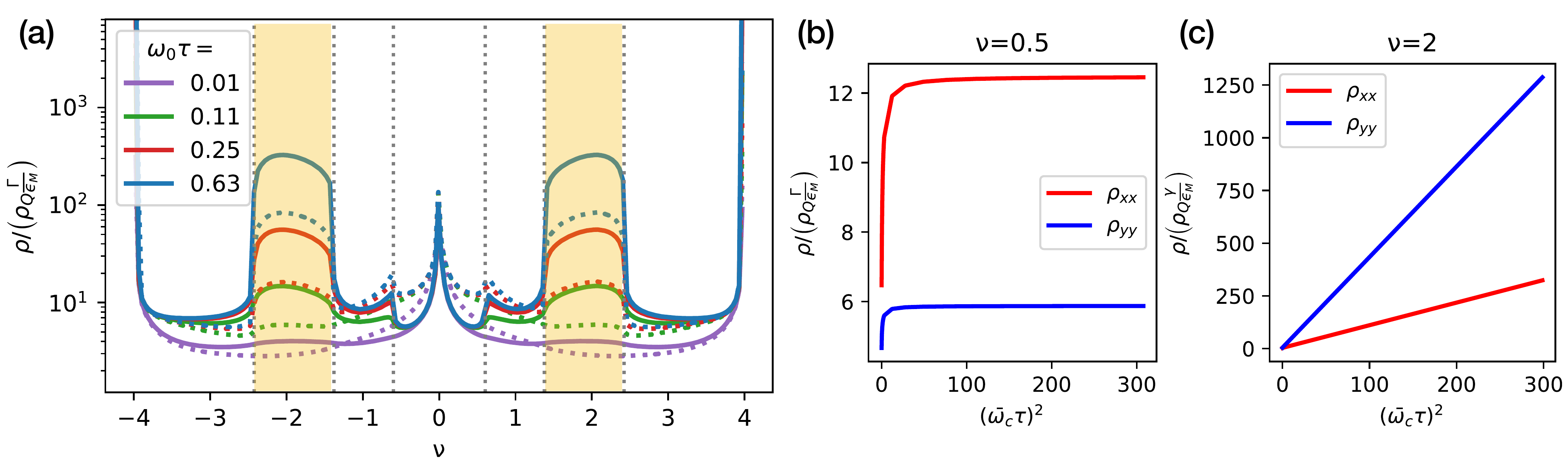}
    \caption{(a) Longitudinal MR $\rho_{xx}$ (dashed) and $\rho_{yy}$ (solid) for strained BM with $\epsilon=0.2\%$ and $\varphi=0^\circ$. Different colors represent varying magnetic field strength. Vertical dashed lines are positions of the van Hove points. Shaded areas are open Fermi surface regions. (b) In the closed Fermi surface region, MR saturates at large magnetic fields. (c) In the open Fermi surface region, MR  exhibit non-saturating $B^2$ dependence along both directions.}
    \label{fig:transport_strained}
\end{figure}

In Fig.~\ref{fig:transport_strained} we show that for the globally defined coordinate system which is misaligned from the principal transport axis, the $B^2$ behavior generally dominates the MR, and therefore will show up in both $\rho_{xx}$ and $\rho_{yy}$ measurements. This remains true for a generic misalignment between the transport axis from experiment and the principal transport axis. 

\begin{figure}
    \centering
    \includegraphics[width=0.45\linewidth]{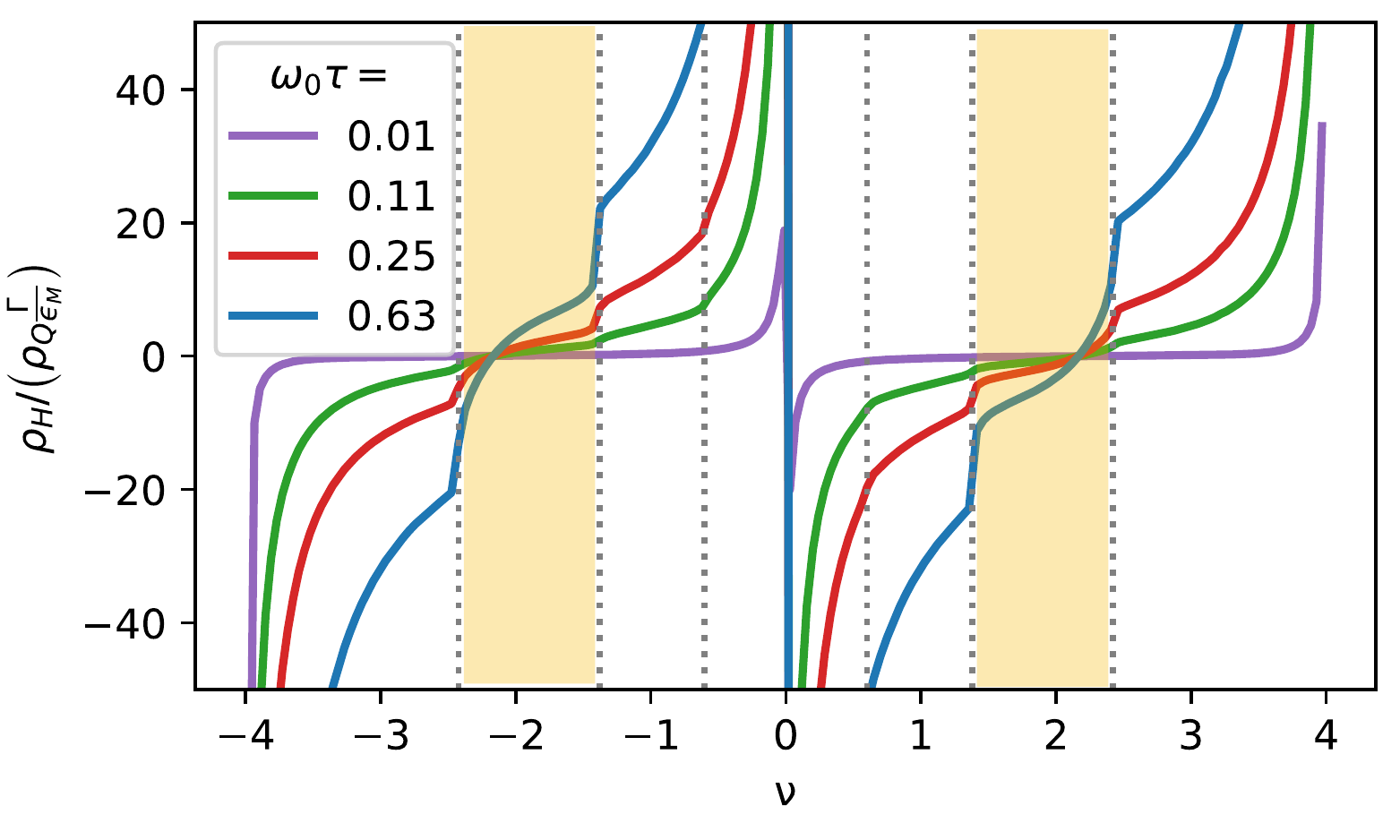}
    \caption{Hall resistivity $\rho_H$ at varying magnetic fields. Note that it crosses zero within the filling range of open Fermi surfaces on both sides of the charge neutrality point. These mark the sign-changing singularities in the Hall number depicted in Fig.~2(b) of the main text.}
    \label{fig:rho21}
\end{figure}

In Fig.~\ref{fig:rho21} we show the Hall resistivity $\rho_H(B)$ for varying magnetic field strength. Since $\rho_H = B/n_Hq$, wherever $\rho_H(B)$ crosses zero and changes sign, the Hall number displays a sign-changing signularity. Fig.~\ref{fig:rho21} clearly shows that $\rho_H(B)$ crosses zero in the open Fermi surface regions on both sides of the charge neutrality point, and independent on the strength of the $B$-field. 

\begin{figure}
    \centering
    \includegraphics[width=0.45\linewidth]{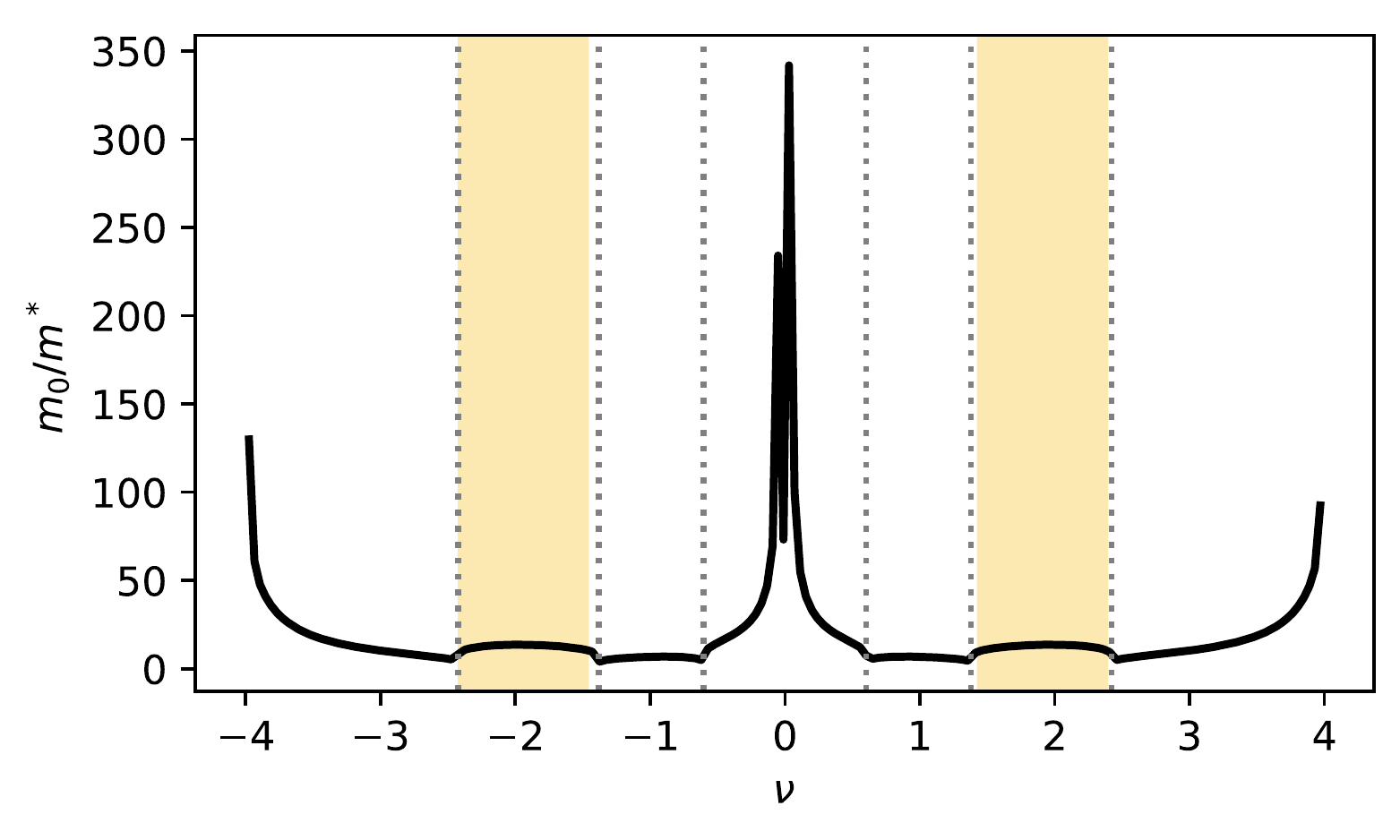}
    \caption{Filling dependence of the averaged inverse cyclotron mass $1/m^*$, extracted from the averaged cyclotron frequency $\bar{\omega_c}=\frac{eB}{m^*}$. Here $\bar{\omega}_c=\frac{1}{N}\sum_{n,i}\omega_{c,n,i}$, where $n$ and $i$ label the FS ($i$) coming from a given band ($n$), in units of the bare electron mass. $1/m^*$ is larger near the charge neutrality and band edges, explaining the earlier onset of quantum oscillations in these filling regions.}
    \label{fig:cyclotron_mass}
\end{figure}

In Fig.~\ref{fig:cyclotron_mass} we illustrate the filling-dependent inverse cyclotron mass $1/m^*$ for strained BM model with $\epsilon=0.2\%$ and $\varphi=0^\circ$. This is to highlight the dichotomy of light-heavy masses on either side of the innermost van Hove singularities closest to the charge neutrality point. This is consistent with the experimental observation of a much earlier onset field of quantum oscillations in filling range below the innermost van Hove point than above.

\begin{figure}
    \centering
    \includegraphics[width=0.9\linewidth]{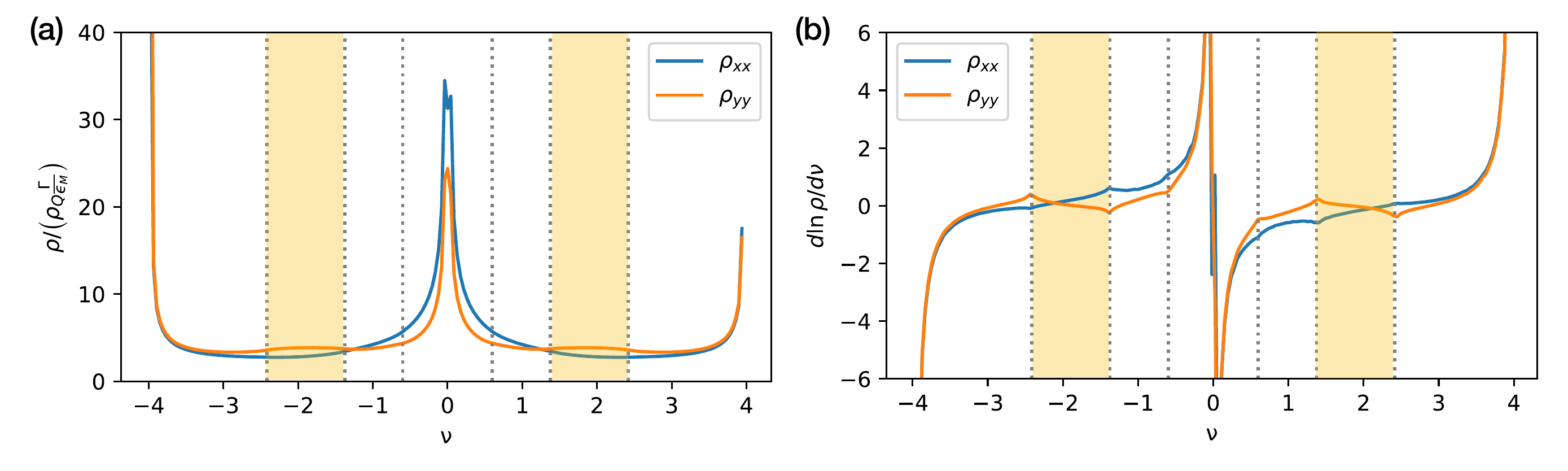}
    \caption{(a) Longitudinal resistivitities $\rho_{xx}$ (blue) and $\rho_{yy}$ (orange) at $B=0$ for strained BM model, with $\epsilon=0.2\%$ and $\varphi=0^\circ$. (b) The derivative of log resistivity with respect to filling. Gray dotted vertical lines mark positions of the van Hove points, and the yellow shaded area marks the open Fermi surface region.}
    \label{fig:transport_B0}
\end{figure}
In Fig.~\ref{fig:transport_B0} we also show the filling dependence of the $B=0$ longitudinal resistivities and their derivatives with respect to filling. A key highlight is that the non-analyticities in the density of states at the van Hove points lead to kink-like features in the derivatives, but nearly invisible in the resistitivies themselves.

\section{Error analysis of antisymmetrization}

\begin{figure}
    \centering
    \includegraphics[width=\linewidth]{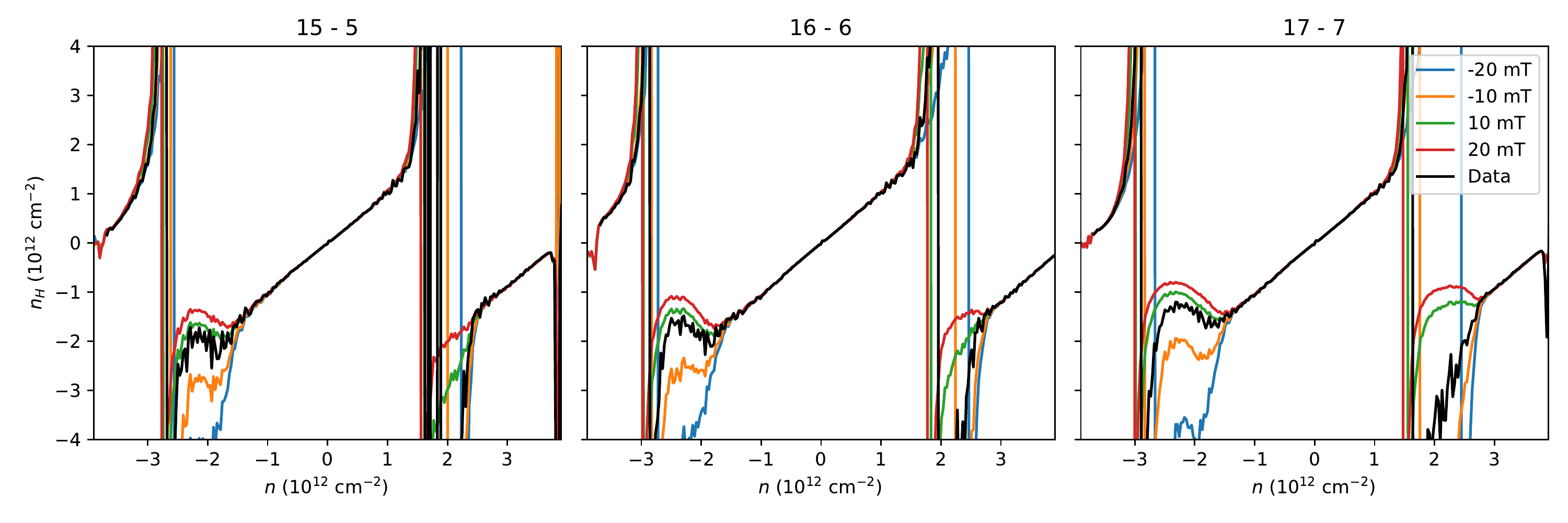}
    \caption{Filling dependent Hall number corrected for antisymmetrization error of the indicated contact pairs at $B = 0.5 \mathrm{T}$. All other transverse contact pairs do not have a large degree of mixing and also do not have these features.}
    \label{fig:error_asm}
\end{figure}
In Fig.~\ref{fig:error_asm} we show that the bump-like features in the experimental Hall number plots in filling range of open Fermi surfaces (Fig.~2(d) of main text and Fig.~\ref{fig:allxy} in the SM) may be attributed to improper antisymmetrization with respect to the $B$-field, namely,
\begin{equation}
  \tilde{\rho}_H(B) = \frac{\rho_{yx}(B+\delta B) - \rho_{yx}(-B+\delta B)}{2},
\end{equation}
where $\delta B$ is a systematic error. The error may be attributed to a small trapped flux of ~10 mT in the superconducting magnet, or perhaps an offset in the magnet power supply. Due to the misalignment of transport principal axis with the Hall bar geometry, longitudinal MR also contributes to $\rho_{yx}(B)$. In the filling range with open Fermi surfaces, the longitudinal resistance exhibits non-saturating quadratic MR, and will mix into the Hall component which is odd in B. As a result, one expects the improper antisymmetrization error to be largest in this filling range.

We investigate this possibility by first fitting a polynomial to the low-field transverse resistivity. This allows us to interpolate the data and add small constant offsets prior to antisymmetrization. Accounting for an offset of roughly 20 mT largely removes the bumps from the data. This offset is larger than what we would expect from trapped flux in a superconducting magnet, however we do not expect the procedure to be accurate to such a fine degree, simply because we do not have fine enough resolution in field to get an accurate polynomial fit.

\section{More experimental measurements based on various contact pairs of the Hall bar geometry}
\begin{figure}
    \centering
    \includegraphics[width=\linewidth]{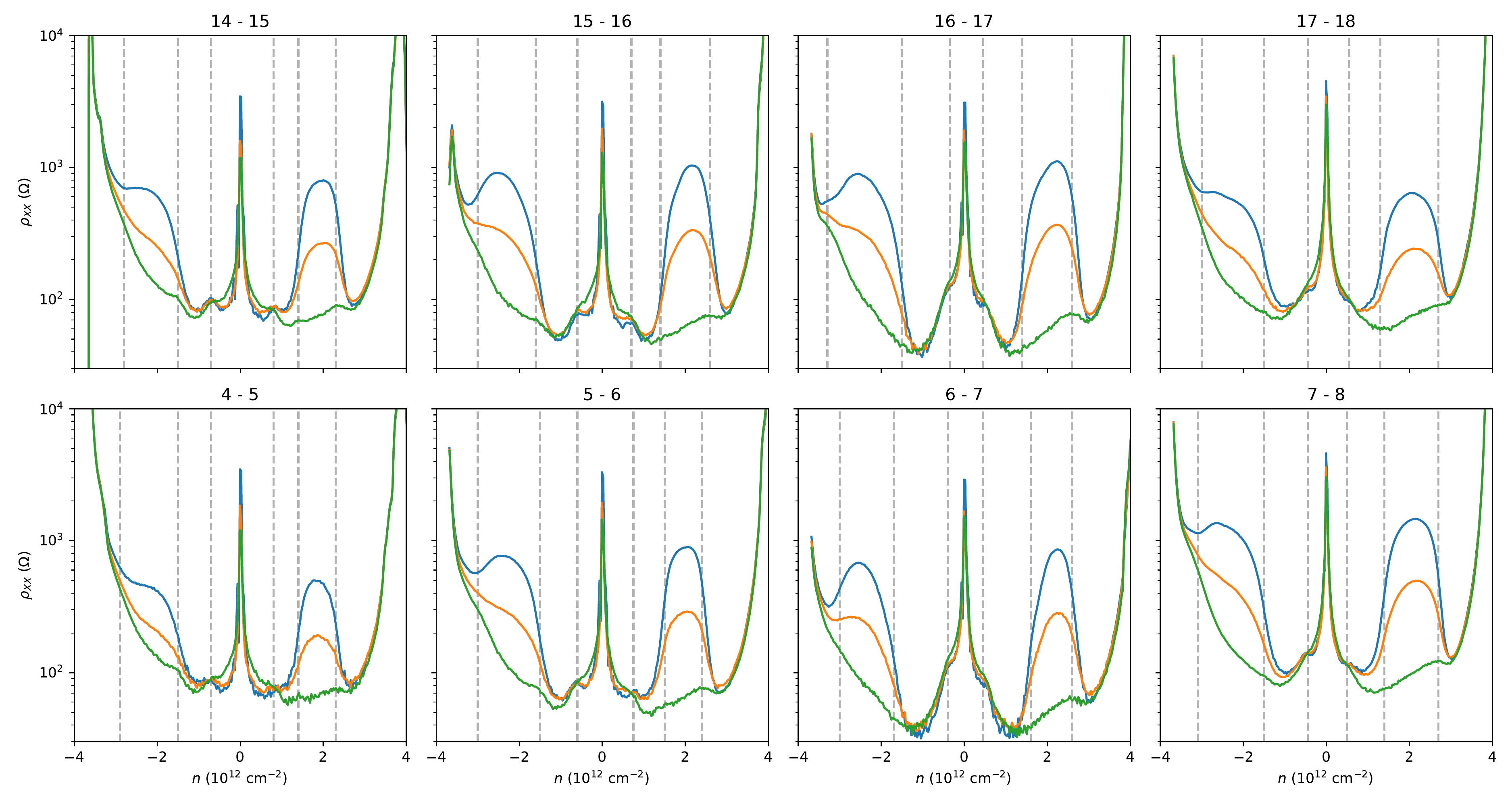}
    \caption{All longitudinal contact pairs with quadratic MR at $B=0$ (green), $0.25$ T (orange), and $0.5$ T (blue), symmetrized. Data taken at 1.6 K. Every contact pair has a well-developed shoulder or cusp near $n/n_s\approx \pm 0.5$ that we associate with the lowest-energy van Hove point. The additional vertical lines are by-eye guesses for the location of the other van Hove points.}
    \label{fig:allxx}
\end{figure}

\begin{figure}
    \centering
    \includegraphics[width=\linewidth]{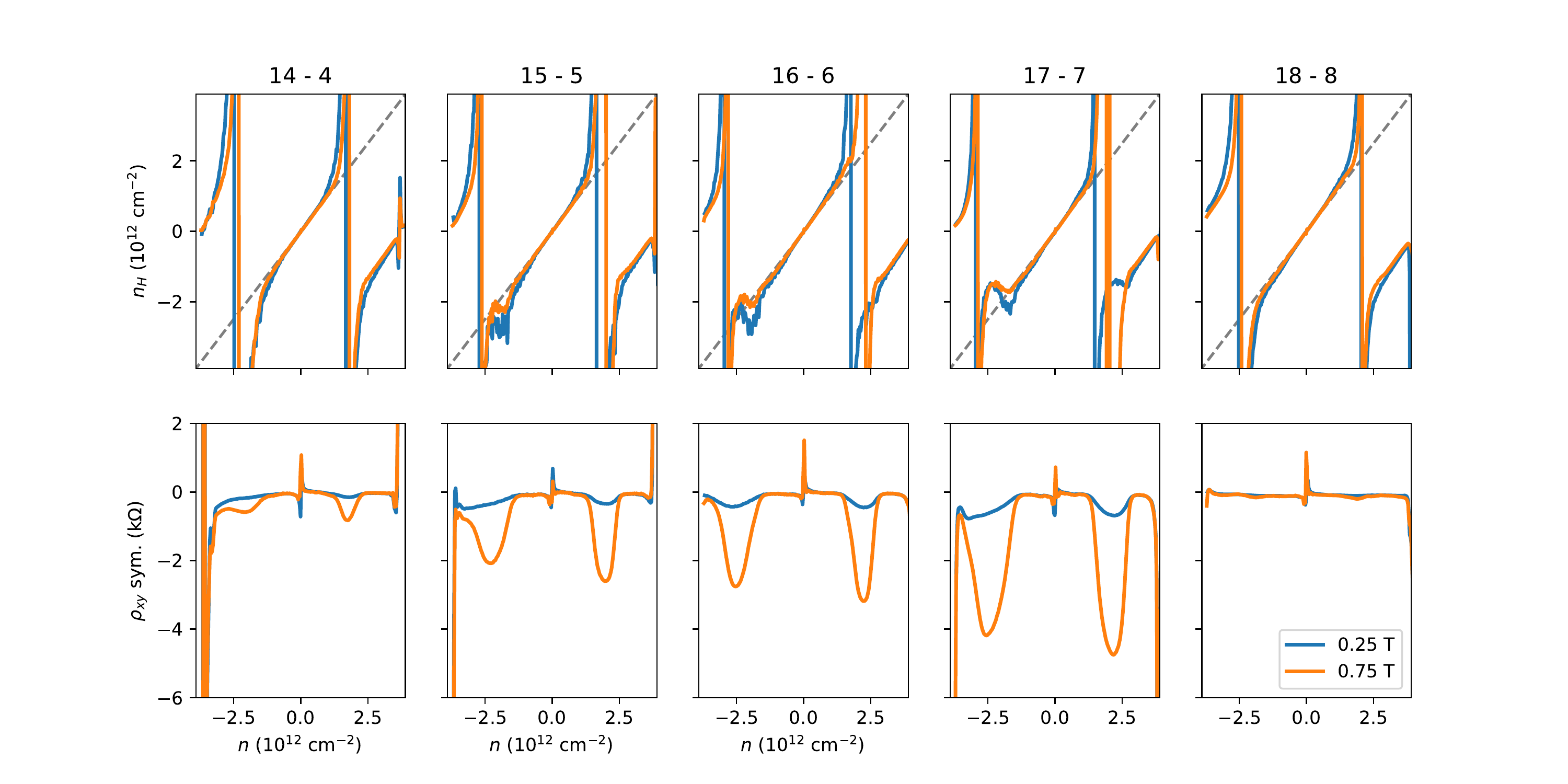}
    \caption{Transverse contact pairs adjacent to the longitudinal pairs presented in Fig.~\ref{fig:allxx}, taken at 1.6 K. Top row: Hall number. Bottom row: symmetrized resistivity. The contact pairs with the largest symmetric component of magnetoresistance display bump-like features in Hall number near where they change sign, consistent with errors in antisymmetrization as illustrated in Fig.~\ref{fig:error_asm}.}
    \label{fig:allxy}
\end{figure}

\begin{figure}
    \centering
    \includegraphics[width=0.8\linewidth]{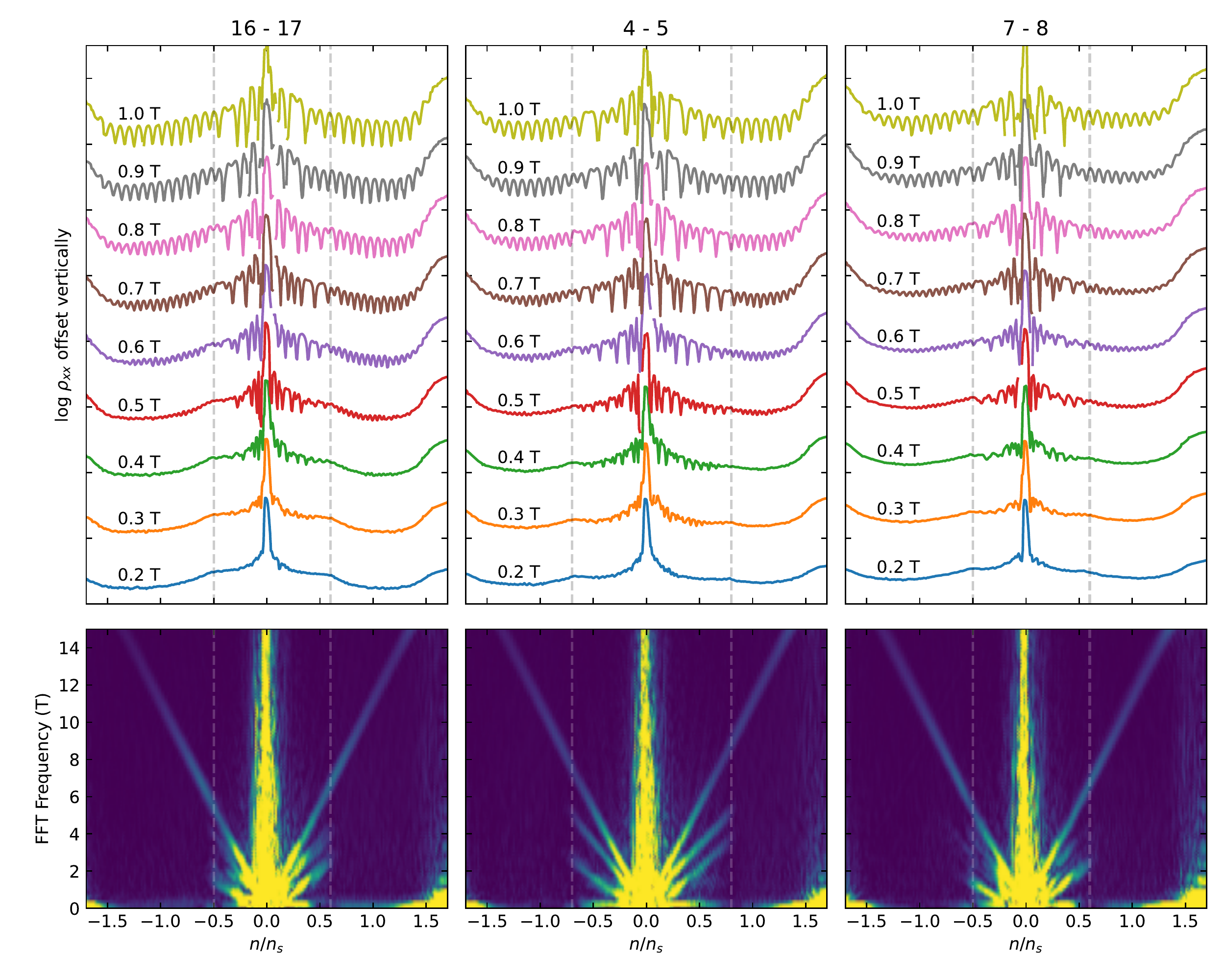}
    \caption{Quantum oscillations near CNP for every longitudinal contact pair for which we have dilution refridgerator data (T = 26 mK). Vertical dashed lines are our estimate of the low-energy van Hove point based on cusps in resistivity at 0.2 T. Contact pair 4 - 5 is the pair shown in the main text.}
    \label{fig:allfftqo}
\end{figure}

The device has nine voltage probes on each side. We observe quadratic magnetoresistance regions in roughly half of the device, between the fourth and eighth contacts. We present longitudinal resistivities of these pairs in Fig.~\ref{fig:allxx}. In each of these pairs, we observe behavior qualitatively consistent with that presented in the main text. Our Hall measurements (Fig.~\ref{fig:allxy}) are similarly consistent.

In Fig.~\ref{fig:allfftqo}, we show quantum oscillations and their Fourier transforms for all three contact pairs for which we have dilution refridgerator data. In all cases, we observe behavior consistent with what we present in the main text: 1) quantum oscillation onset at lower field close to CNP, 2) an irregular pattern of resistivity minima close to CNP, and 3) extra features in the FFT of the quantum oscillations that end at vH1. The density of the first van Hove point is closer to the CNP in the other two contact pairs, and the extra features in the FFT are not as clear.

\end{document}